\documentclass[epj,nopacs]{svjour}
\usepackage[utf8]{inputenc}
\usepackage{amsmath,amssymb,amsfonts,graphicx,cite,multirow}
\usepackage{tikz}
\usepackage{enumitem}

\usepackage{pstricks,pdflscape,morefloats}
\usepackage{booktabs}
\usepackage{appendix}
\usepackage[section]{algorithm}
\usepackage{algorithmicx}
\usepackage{algpseudocode,slashed}

\graphicspath{{graphics/}}
\makeatletter
\ifx\input@path\@undefined
\def\input@path{{graphics/}}
\else
\g@addto@macro\input@path{{graphics/}}
\fi
\makeatother

\usepackage{color}
\usepackage{hyperref}

\usepackage{mathtools}

\preprint{KA-TP-32-2017\\UWTHPH-2017-34\\HERWIG-2017-03\\MCNET-17-25}

\title{Baryon production from cluster hadronization}

\author{Stefan Gieseke\inst{1}, Patrick Kirchgae\ss{}er\inst{1}, Simon Pl\"atzer\inst{2} }

\institute{Institute for Theoretical Physics, Karlsruhe Institute of
  Technology, 76128 Karlsruhe, Germany \and Particle Physics, Faculty of
  Physics, University of Vienna, 1090 Vienna, Austria}

\date{\today}

\abstract{We present an extension to the colour reconnection model in the
  Monte-Carlo event generator Herwig to account for the production of baryons
  and compare it to a series of observables for soft physics.  The new model
  is able to improve the description of charged-particle mutliplicities and
  hadron flavour observables in pp collisions.  \PACS{{xx.yy.zz}{Xx Yy Zz}} }

\hypersetup{draft}
\begin{document}

\maketitle

\section{Introduction}
With increasing precision from the LHC it becomes apparent that many
non-perturbative aspects of elementary particle production are far from
understood.  Especially the description of the transition from the
deconfined state to final state particles that are observed in the
detectors has many unknown variables and raises a lot of questions.
With the help of Monte-Carlo event generators\,
\cite{Bahr:2008pv,Bellm:2015jjp,Bellm:2017bvx,Pythia8.2,Sherpa} different models can
be evaluated.  Among the problems, that are being observed are the
correct description of high-multiplicity events and the flavour
composition of final states.  One striking observation, made recently by
the \mbox{ALICE} collaboration showed that in high-multiplicity pp
events, properties similar to that of AA and pA collisions are
observed\,\cite{ALICE:2017jyt}.

Possible explanations of these effects are rooted in the possibility
that partonic matter shows some collective behaviour as in a
hydrodynamical description, see e.g.\ \cite{Werner:2013tya}.  The other
route to introduce strong and possibly quite long-range correlations
among different hard partons in a single interaction goes via colour
reconnections.  Here, states of high partonic density may lead to some
kind of absorption or neutralization of colour charge.  These ideas have
been advocated in some way e.g.\ in the Dipsy rope
model\,\cite{Bierlich:2014xba} where many overlapping strings are
combined into a colour field of a higher representation.
Thermodynamical string fragmentation in Pythia also addresses this issue
where shifts of the transverse momentum of heavier particles to higher
values are the main result\,\cite{Fischer:2016zzs}.  The possibility to
form string junctions within the Lund string fragmentation model has
been introduced in \cite{Christiansen:2015yqa}.

In Herwig an accurate description of Minimum Bias (MB) and Underlying Event
(UE) observables has been achieved with the recent development of a new model
for soft and diffractive interactions\,\cite{Gieseke:2016fpz}, building
on the earlier developments 
in\,\cite{Bahr:2008wk,Bahr:2008dy,Borozan:2002fk,Butterworth:1996zw}.  Here,
the importance of colour reconnections has already been observed.
However, in this work only charged particles have been addressed as such
and we have already pointed out shortcomings in the description of high
multiplicity tails.  This observation lead to the consideration that the
mere production of baryons by itself would lead to a reduction of
charged multiplicity in favour of a rise of the multiplicity of heavier
particles.  We do not address effects that arise at high multiplicity in
particular but rather aim for an improved global description of particle
production in MB events.  

In this study we therefore introduce a possible extension to the model
for colour reconnection to account for the production of baryons.  At
the same time we reconsider the production of strange particles and find
that with a slight modification of our parameters we can improve the
production rates of strange mesons as well as baryons quite
significantly.  We compare these effects to recent observations made by
CMS and ALICE.  Especially charged-particle multiplicities and ratios of
identified hadrons are of main interest.

\section{Colour reconnection}
\label{sec:colourreconnection}
In order to describe the full structure of a particle scattering process
additional soft effects that are not accessible by perturbation theory
have to be considered.  Such effects include hadronization, Multiple
Parton Interactions (MPI) and fragmentation processes.  In general these
non-perturbative effects are based on phenomenological considerations.
The basis for the hadronization model in Herwig is the cluster
model\,\cite{WEBBER1984492}, which forms colourless singlets from colour
connected partons.  The fragmentation of these clusters into hadrons
depends on the invariant cluster mass and the flavour of the quarks
inside the cluster.  The colour connections between the partons in an
event are determined by the $N_C\rightarrow \infty$ approximation which
leads to a planar representation of colour lines\,\cite{HOOFT1974461}.
Every quark is connected to an antiquark and gluons, carrying both
colour and anticolour are connected to two other partons.  The goal of
colour reconnection is to study whether different connection topologies,
other than the predefined colour connection are possible between the
partons.

In hadronic collisions the colour reconnection mostly aims at a
resurrection of the colour correlation between different hard partonic
interactions.  Within the Monte Carlo modeling of MPI, 
different hard partonic scatters are layered on top
of each other without a clear understanding of how to introduce a
pre-confined state when co-moving partons from different scattering
centers should also lead to `closeness' in colour space, i.e.\ to short
colour lines between those partons.  The importance of the effect has
first been observed in \cite{Sjostrand:1987su}.  The colour reconnection
leads to a decrease of charged multiplicity for a given partonic
configuration and hence an increase of the average transverse momentum
per charged particle.  The effect gets stronger with denser states,
e.g.\ as we increase the CM energy of the hadronic collider.

The effects of colour reconnection have also been studied in the context of
$\rm{W}^+\rm{W}^-$ production at
LEP-2\,\cite{GUSTAFSON198890,Sjostrand:1993hi}.  Due to the large space-time
overlap of the decaying bosons the two hadronic systems may be in contact with
each other which leads to colour interchange and can cause one quark of the
$\rm{W}^+$ boson to hadronize together with an antiquark of the $\rm{W}^-$
boson.

\subsection{The colour reconnection model in Herwig}

\begin{figure}[t]
\centering
\includegraphics[width=8cm]{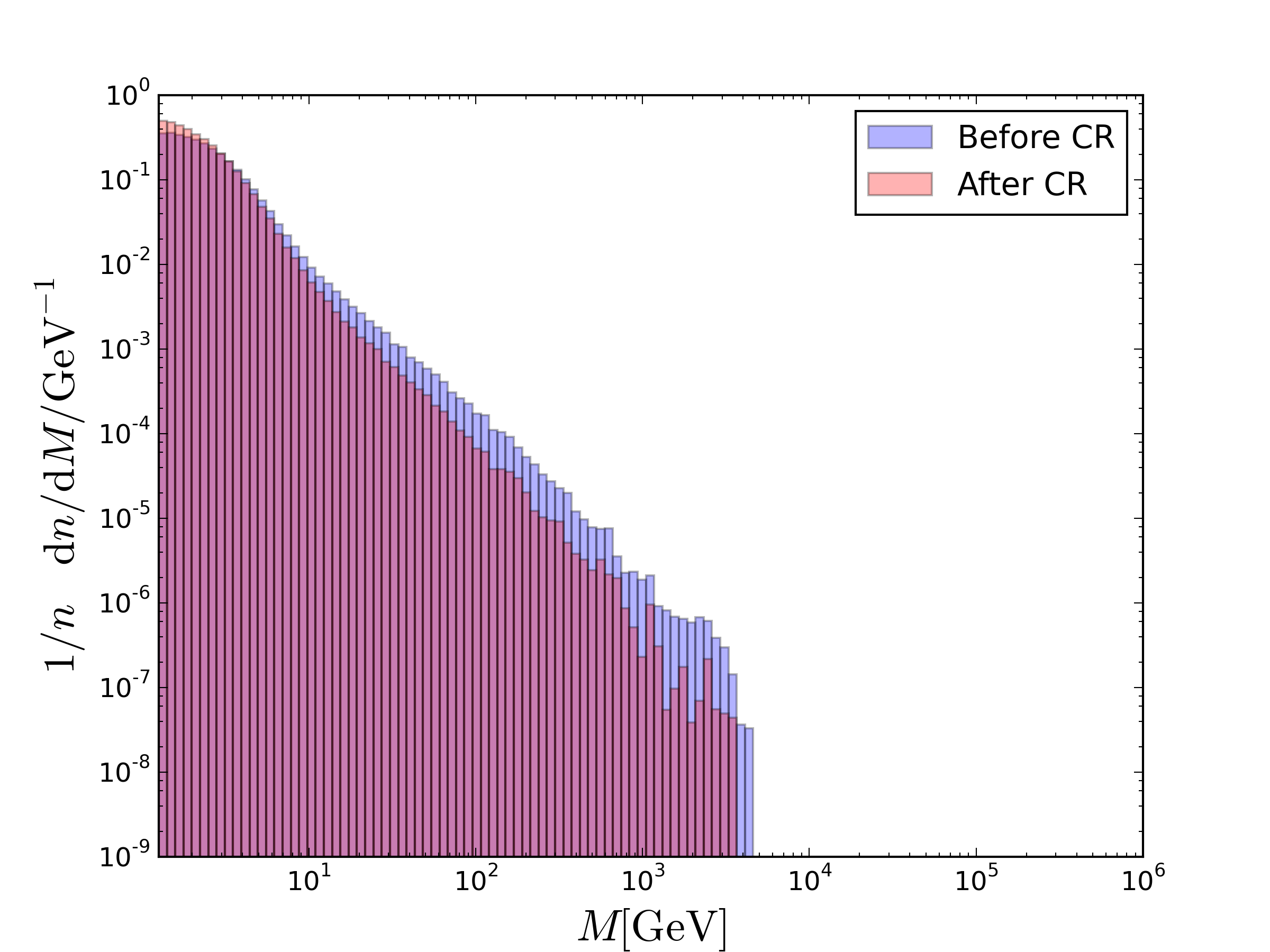}
\caption{Distribution of invariant cluster masses before and after colour reconnection 
	for a pp collision with centre-of-mass energy of $\sqrt{s}=7\,\mathrm{TeV}$}. 
\label{fig:defaultmass}
\end{figure}

The algorithm for colour reconnection in Herwig is implemented directly before
the cluster fission takes place\,\cite{roehr:phd}.  The properties of a
cluster are defined by the invariant cluster mass
\begin{align}
 M^2 = (p_1 + p_2)^2,
\end{align}
where $p_1$ and $p_2$ are the four momenta of the cluster constituents.  The
fission and the decay of the cluster depend on the invariant cluster mass
which directly influences the multiplicity of final state particles.  Two
algorithms for colour reconnection are currently implemented in Herwig, the
plain colour reconnection and the statistical colour
reconnection\,\cite{roehr:phd}.  Both algorithms try to find configurations of
clusters that would reduce the sum of invariant cluster masses,
\begin{align}
\lambda = \sum_{i=1}^{N_{\mathrm{cl}}}M_{i}^2, 
\end{align}
where $N_{\mathrm{cl}}$ is the number of clusters in an event. 
The plain colour reconnection
algorithm picks a cluster randomly from the list of clusters and compares it
to all other clusters of that list. For every cluster the invariant masses of
the original cluster configuration $M_{\rm{A}}+M_{\rm{B}}$ and the masses of
the possible new clusters $M_{\rm{C}} + M_{\rm{D}}$ are calculated. The
cluster configuration that results in the lowest sum of invariant cluster
masses is then accepted for reconnection with a certain probability
$p_{\rm{R}}$. If the reconnection is accepted the clusters $(\rm{A})$ and
$(\rm{B})$ are replaced by the clusters $(\rm{C})$ and $(\rm{D})$. This
algorithm works out clusters with lower invariant masses and therefore
replaces heavier clusters by lighter ones.
In Fig.\ref{fig:defaultmass} we show the distribution 
of invariant cluster masses for a centre-of-mass energy of $\sqrt{s}=7\,\mathrm{TeV}$
before and after the plain colour reconnection. Due to the colour reconnection 
procedure the distribution is enhanced in the low-mass peak region and suppressed in 
the high-mass tail.

 The statistical colour
reconnection on the other hand uses a simulated annealing algorithm to find
the configuration of clusters that results in the absolute lowest value of the
colour length $\lambda$.  While being computing intensive it was also found
in\,\cite{Gieseke:2012ft} that the statistical colour reconnection prefers a
quick cooling that does not result in a global minimum of colour length
$\lambda$ in order to describe the data best. 
In a recent paper the colour reconnection model was
changed in a way, that it is forbidden to make a reconnection which
would lead to a gluon produced in any stage of the parton-shower evolution
becoming a colour-singlet after hadronization\,\cite{Reichelt:2017hts}.

\subsection{Extension to the colour reconnection model}

The only constraint upon forming a cluster is that the cluster has to be able
to form a colourless singlet under $SU(3)_C$. In $SU(3)_C$ a coloured quark is
represented as a triplet (3) and an anticoloured antiquark is represented as
an anti-triplet $(\bar{3})$.  Two triplets can be represented as an
anti-triplet and two anti-triplets can be represented as a triplet,
\begin{align}
\label{CR:eq1}
  3 \otimes 3 &= 6 \oplus \bar{3},\\
  \bar{3} \otimes \bar{3} &= \bar{6} \oplus 3.
\end{align}
The clusters are a combination of these coloured quarks were only combinations
are allowed that result in a colourless singlet.  Here we consider the
following allowed cluster configurations based on the $SU(3)_C$ structure of
QCD.  We begin with the normal cluster configuration which will be referred to
as a \textit{mesonic} cluster
\begin{align}
 3 \otimes \bar{3} = 8 \oplus 1.
\end{align}
In strict $SU(3)_C$ the probability of two quarks having the correct colours
to form a singlet would be 1/9.  Next we consider possible extensions to the
colour reconnection that allows us to form clusters made out of 3 quarks. A
\textit{baryonic} cluster consists of three quarks or three antiquarks where
the possible representations are,
\begin{align}
 3 \otimes 3 \otimes 3 &= 10 \oplus 8 \oplus 8 \oplus 1,\\
 \bar{3} \otimes \bar{3} \otimes \bar{3} &= 10 \oplus 8 \oplus 8 \oplus 1.
\end{align}
In full $SU(3)_C$ the probability to form a singlet made out of three quarks
would be 1/27.  In the following we will introduce the algorithm we used for
the alternative colour reconnection model.  In order to extend the current
colour reconnection model, which only deals with \textit{mesonic} clusters, we
allow the reconnection algorithm to find configurations that would result in a
\textit{baryonic} cluster.

\subsection{Algorithm}

\begin{figure}[t]
\centering
\includegraphics[width=7cm]{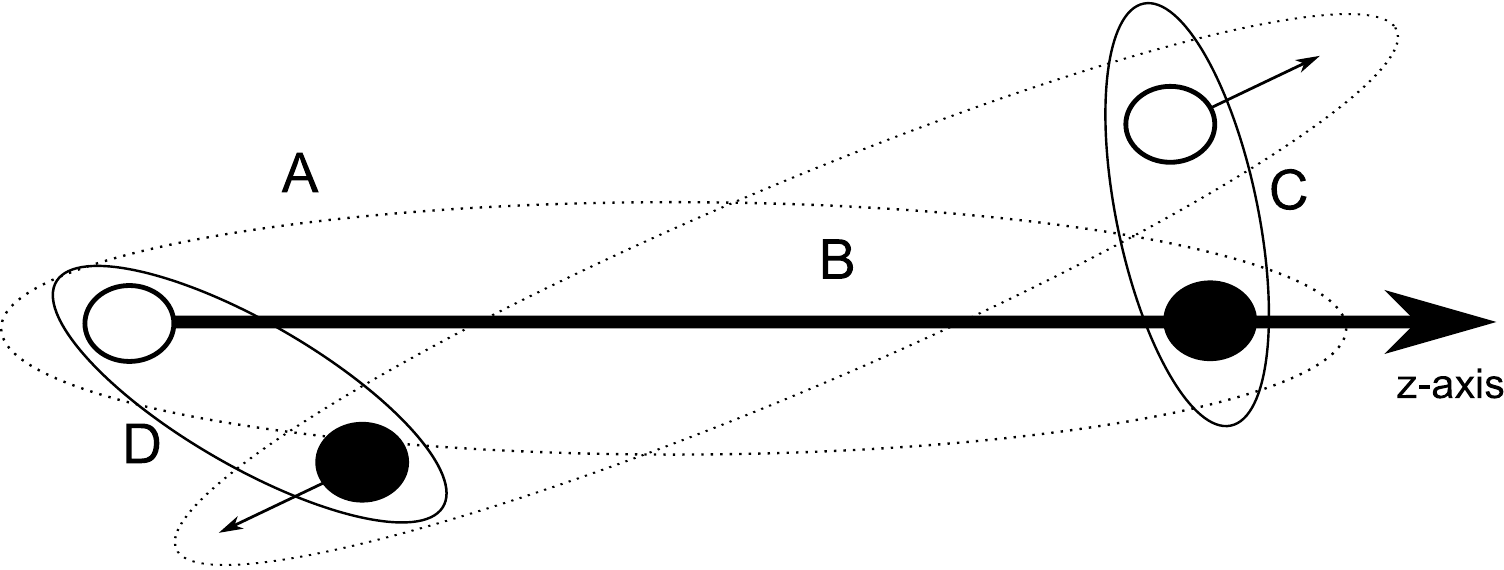}
\caption{ Representation of rapidity based colour reconnection where the quark
  axis of one cluster is defined as the z-axis in respect to which the
  rapidities of the constituents from the possible reconnection candidate are
  calculated. $(\rm{A})$ and $(\rm{B})$ are the the original
  clusters. $(\rm{C})$ and $(\rm{D})$ would be the new clusters after the
  reconnection.  }
\label{fig:rapidityCR}
\end{figure}

\begin{figure}[t]
\centering
\includegraphics[width=7cm]{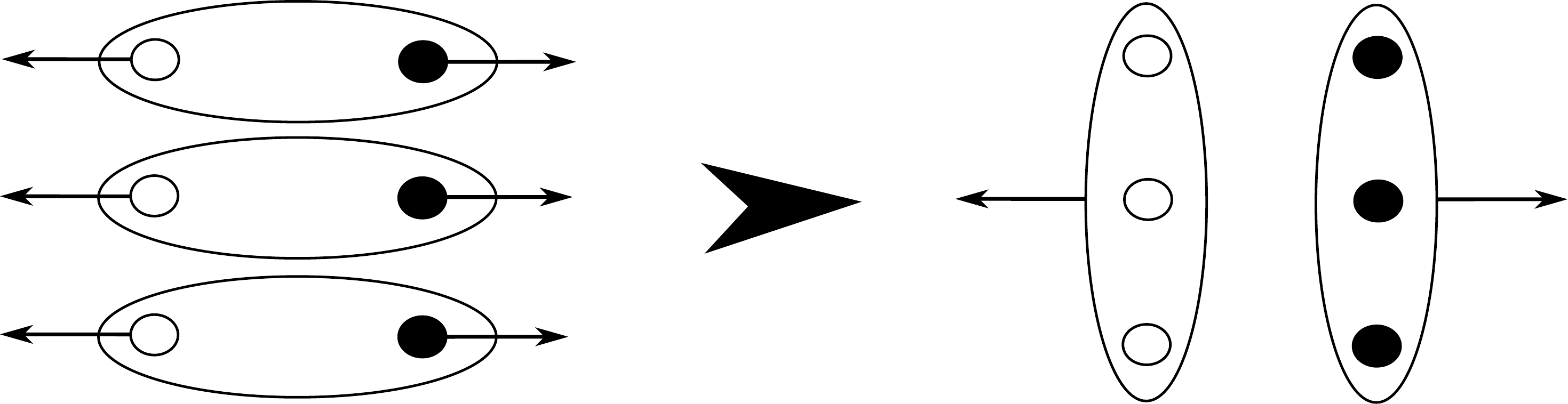}
\caption{ Configuration of clusters that might lead to baryonic reconnection.
  The small black arrows indicate the direction of the quarks.  A reconnection is
  considered if all quarks move in the same direction and all antiquarks move
  in the same direction.  }
\label{fig:baryonic}
\end{figure}

\begin{figure}[t]
\centering
\includegraphics[width=9cm]{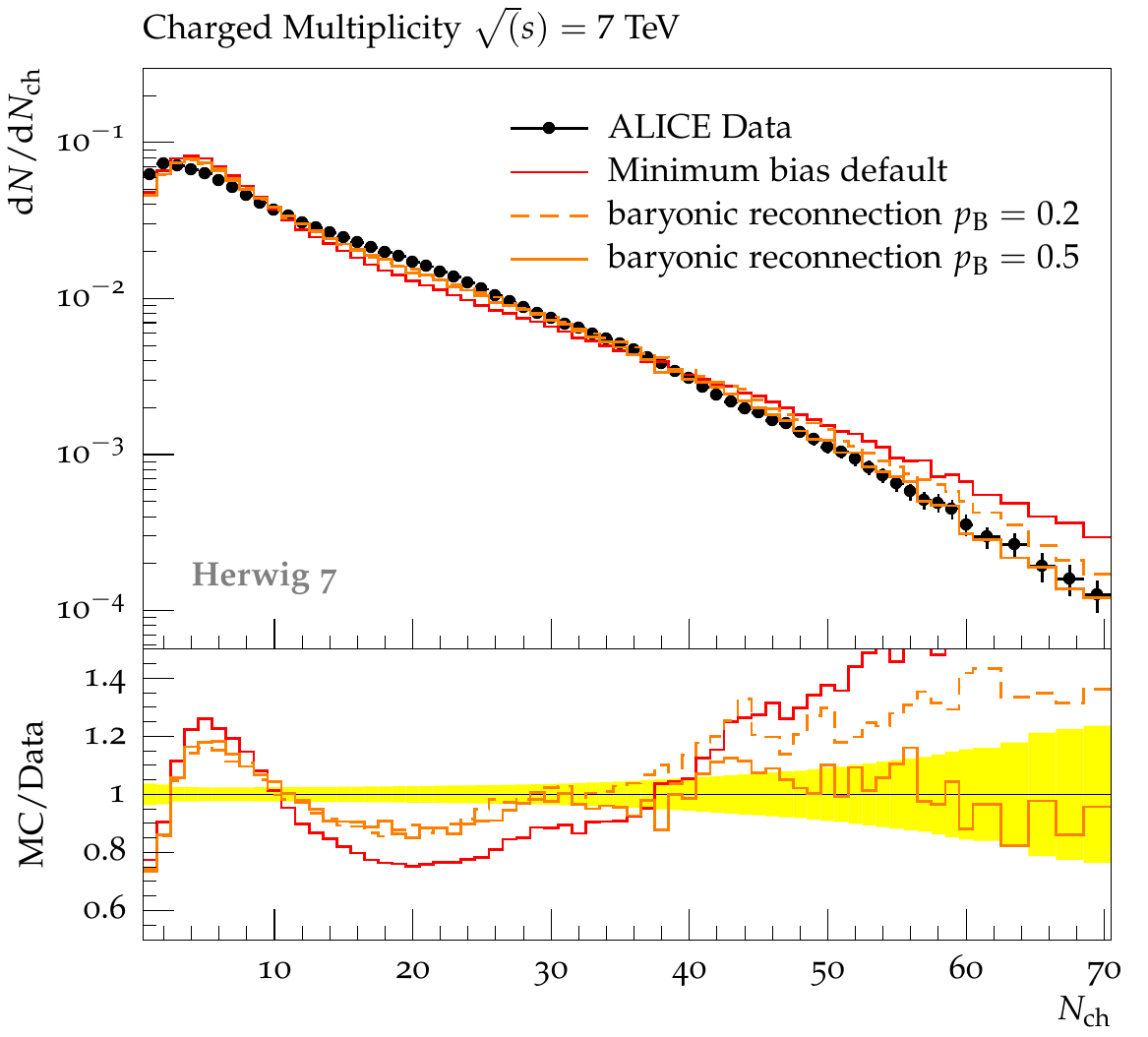}
\caption{ Measurement of the charged-particle multiplicity at
  $\sqrt{s}=7\,\mathrm{TeV}$ with ALICE at LHC\,\cite{Aamodt:2010pp}. Shown is
  a comparison of the new colour reconnection model for different reconnection
  probabilities with the default model of Herwig.  }
\label{fig:nch}
\end{figure}

\begin{figure*}[t]
  \centering
  \includegraphics[width=0.49\textwidth]{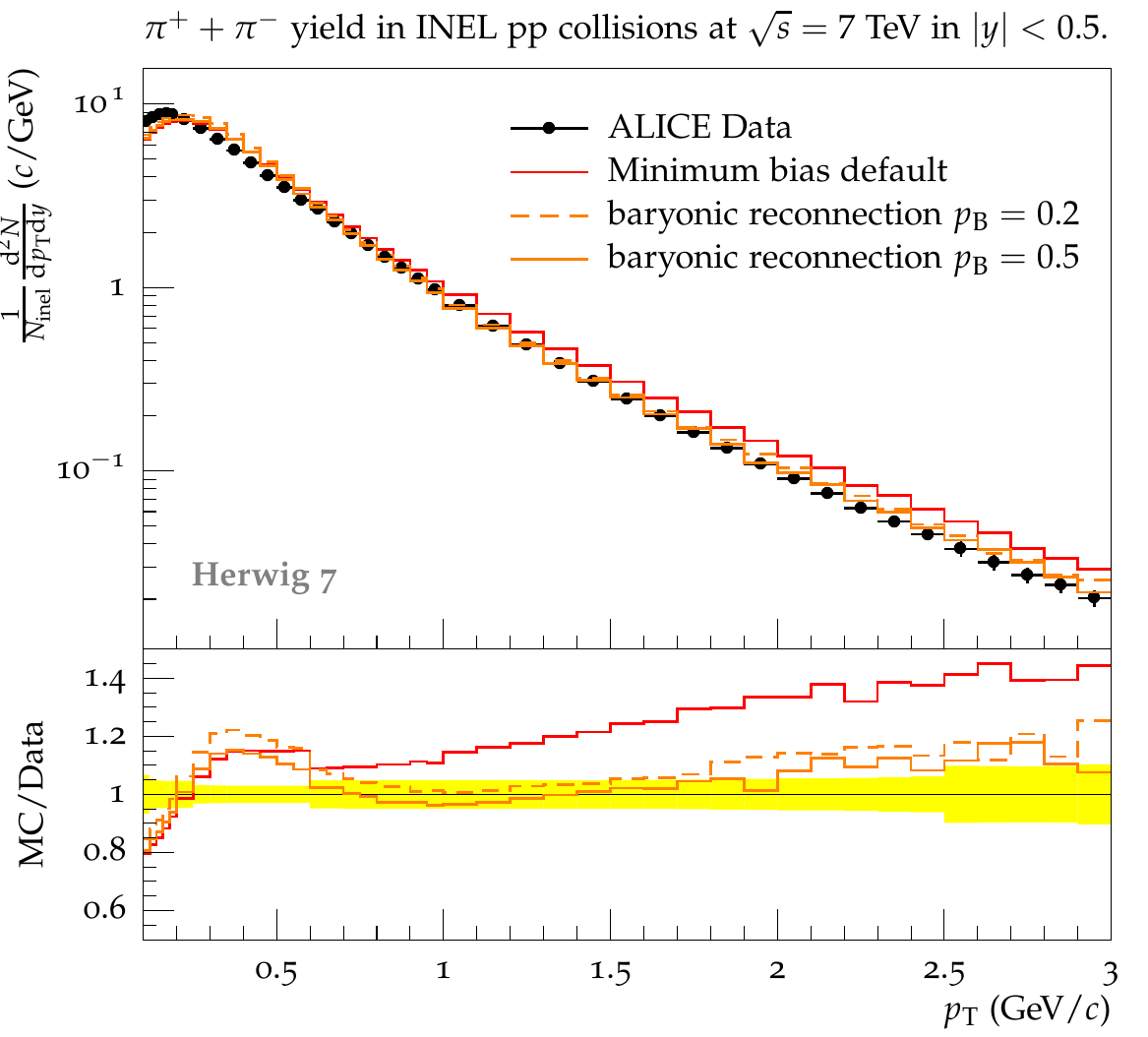}\hfill
  \includegraphics[width=0.49\textwidth]{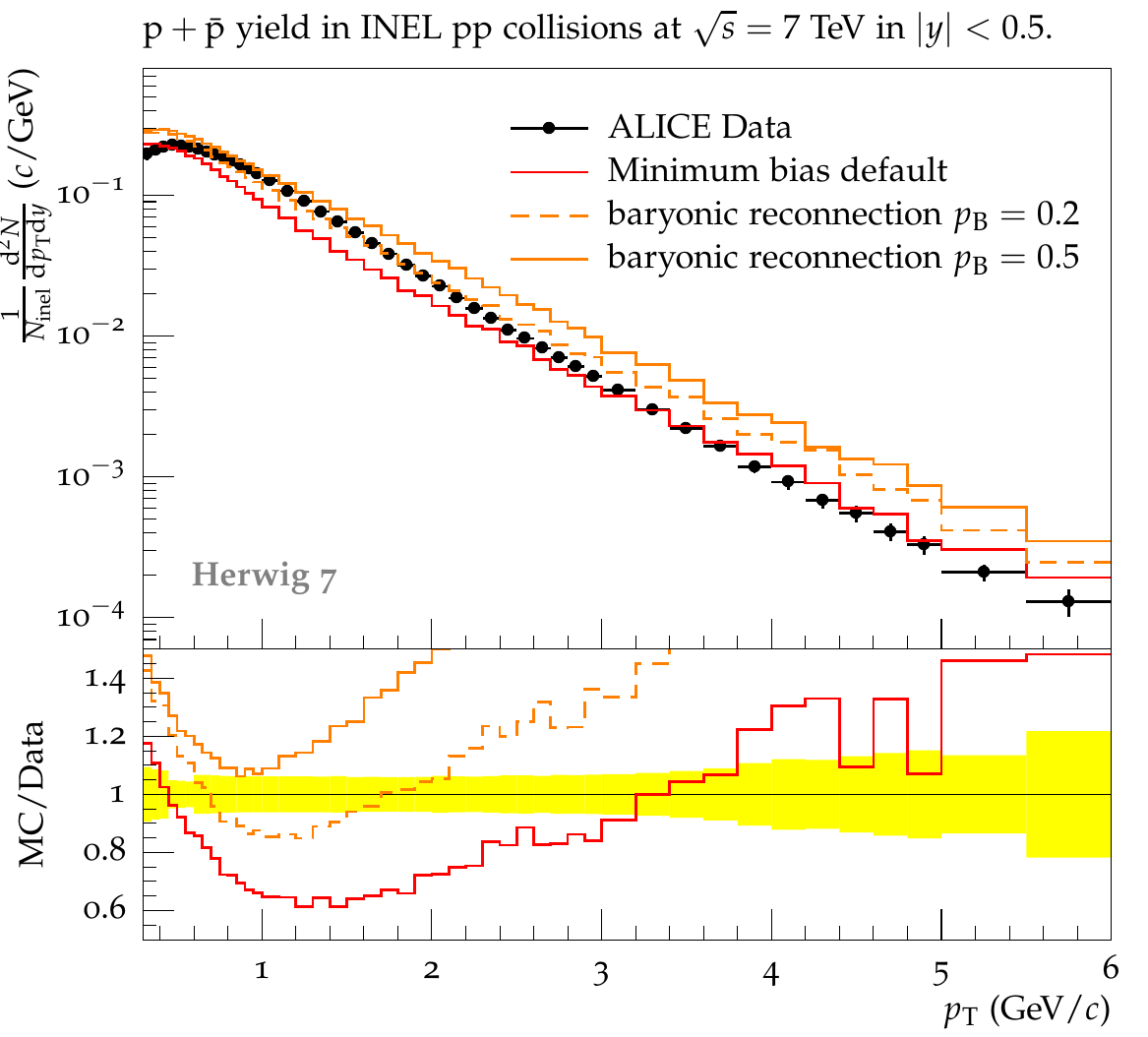}
  \caption{The transverse momentum spectra for $\mathrm{\pi^++\pi^-}$ and $\mathrm{K^++K^-}$ as measured by
            ALICE at $\sqrt{s}=7\,\mathrm{TeV}$ \cite{Adam:2015qaa} in the very central rapidity region $|y|<0.5$.
    }
  \label{fig:baryonicCR}
\end{figure*}

As explained before the colour reconnection algorithms in Herwig are
implemented in such a way that they lower the sum of invariant cluster masses.
For baryonic reconnection such a condition is no longer reasonable because of
the larger invariant cluster mass a baryonic cluster carries.  As an
alternative we consider a simple geometric picture of nearest neighbours were
we try to find quarks that approximately populate the same phase space region
based on their rapidity $y$.  The rapidity $y$ is defined as
\begin{align}
y = \frac{1}{2}\ln \left(\frac{E+p_z}{E-p_z}\right),
\end{align}
and is usually calculated with respect to the $z$-axis.  Here we consider
baryonic reconnection if the quarks and the antiquarks are flying in the same
direction.  This reconnection forms two baryonic clusters out of three mesonic
ones.  The starting point for the new rapidity based algorithm is the
predefined colour configuration that emerges once all the perturbative
evolution by the parton shower has finished and the remaining gluons are split
non-perturbative\-ly into quark-antiquark pairs.  Then a list of clusters is
created from all colour connected quarks and anti-quarks.  The final algorithm
consists of the following steps:
\begin{enumerate}
	\item Shuffle the list of clusters in order to prevent the bias that comes from 
		the order in which we consider the clusters for reconnection
	\item Pick a cluster $(\rm{A})$ from that list and boost into the rest-frame of that 
		cluster. The two constituents of the cluster $(q_{\rm{A}}, \bar{q}_{\rm{A}})$ 
		are now flying back to back and we define the direction of the 
		antiquark as the positive $z$-direction of the quark axis.
	\item Perform a loop over all remaining clusters and calculate the rapidity of the
		cluster constituents with respect to the quark axis in the rest frame of
		the original cluster for each other cluster in that list $(\rm{B})$.
	\item Depending on the rapidities the constituents of the cluster $(q_{\rm{B}}, \bar{q}_{\rm{B}})$ 
		fall into one of three categories:
	\begin{itemize}
        	\item[] Mesonic: $y(q_{\rm{B}}) > 0 > y(\bar{q}_{\rm{B}})~.$
        	\item[] Baryonic: $y(\bar{q}_{\rm{B}}) > 0 > y(q_{\rm{B}})~.$
		\item[] Neither.
	\end{itemize}
		If the cluster neither falls into the mesonic, 
		nor in the baryonic category listed above 
		the cluster is not considered for reconnection.
	\item The category and the absolute value $|y(q_{\rm{B}})| +|y(\bar{q}_{\rm{B}})|$ 
		for the clusters with the two largest sums is saved (these are clusters B 
 		and C in the following).
	\item Consider the clusters for reconnection depending on their category. 
		If the two clusters with the largest sum (B and C) are in the category 
		\textit{baryonic} consider them for baryonic reconnection (to cluster A) with 
		probability $p_{\rm{B}}$. If the category of the cluster with the largest sum 
		is \textit{mesonic} then consider it for normal reconnection with probability 
		$p_{\rm{R}}$. If a baryonic reconnection occurs, 
		remove these clusters (A, B, C) from the list and do not consider them 
		for further reconnection.
		A picture of the rapidity based reconnection for a mesonic configuration is
		shown in Fig. \ref{fig:rapidityCR} and a simplified sketch for baryonic 
		reconnection is shown in Fig. \ref{fig:baryonic}.
	\item Repeat these steps with the next cluster in the list.
\end{enumerate}
We note that with this description we potentially exclude clusters from
reconnection where both constituents have a configuration like $y(q_{\rm{B}})
> y(\bar{q}_{\rm{B}}) > 0$ w.r.t.\ the quark axis but assume that these clusters
already contain constituents who are close in rapidity and fly in the same
direction.  The exclusion of baryonically reconnected clusters from further
re-reconnection biases the algorithm towards the creation of baryonic clusters
whose constituents are not the overall nearest neighbours in rapidity.  The
extension to the colour reconnection model gives Herwig an additional
possibility to produce baryons on a different, more elementary level than on
the level of cluster fission and cluster decay\,\cite{Bahr:2008pv}.  In pp
collisions with enhanced activity from MPI a high density of clusters leads to
an increased probability of finding clusters that are suitable for baryonic
reconnection.  We expect this model therefore to have a significant effect on
charged-hadron multiplicities, especially on the high-multiplicity region.  We
also expect the new model to have a significant impact on baryon and meson
production since baryonic colour reconnection effectively makes baryons out of
mesons.  In Figs. \ref{fig:nch} and \ref{fig:baryonicCR} we see the influence
of the new model for different values of $p_{\mathrm{B}}$ on the
charged-particle multiplicities and the $p_{\perp}$ spectra of $\pi^++\pi^-$
and $\rm{p}+\bar{\rm{p}}$ yields in inelastic pp collisions at $\sqrt{s} = 7
\,\mathrm{TeV}$ in the central rapidity region.  As expected the model
influences the hadronic multiplicities for large $N_{\rm{ch}}$ significantly.
A larger baryonic reconnection probability reduces the number of high
multiplicity events and shifts them towards lower multiplicities.  The
$p_{\perp}$ distribution of the $\pi^++\pi^-$ shows an overall reduction while
the $p_{\perp}$ spectra of the $\rm{p}+\bar{\rm{p}}$ shows an overall
enhancement due to baryonic colour reconnection.  While the description of the
low $p_{\perp}$ region improves, there are too many $\rm{p}+\bar{\rm{p}}$ with
a $p_{\perp} > 2.5\,\rm{GeV}$.  In the next section we describe the tuning of
the model to a wide range of data from hadron colliders.

\section{Tuning}
\label{sec:tuning}

The tuning is achieved by using the Rivet and Professor framework for Monte-Carlo event
generators\,\cite{Buckley:2010ar,Buckley:2009bj}.  In a first tuning attempt we keep
the hadronization parameters that were tuned to LEP data at their default
values and follow a similar tuning procedure as in\,\cite{Gieseke:2016fpz}.
We retune the main parameters of the MPI model in Herwig, the
$p_{\perp,0}^{\rm{min}}$ parameter and the inverse proton radius squared
$\mu^2$.  Since we altered the colour reconnection model, we also retune the
probability for normal colour reconnection $p_{\mathrm{R}}$.  The only
additional parameter we have to consider is the probability for baryonic
reconnection $p_{\mathrm{B}}$.  In order to capture general features of MB
observables we tune the model to a large variety of MB data from the ATLAS and
ALICE collaborations at
$\sqrt{s}=7\,\rm{TeV}$\,\cite{Adam:2015qaa,Aad:2010ac}.  The following
observables were used with equal weights:
\begin{itemize}
	\item[•] The pseudorapidity distributions for\\ $N_{\rm{ch}}\geq 1$, $N_{\rm{ch}}\geq 2$, $N_{\rm{ch}}\geq 6$, $N_{\rm{ch}}\geq 20$,
	\item[•] The transverse momentum of charged particles for \\$N_{\rm{ch}}\geq 1$,
	\item[•] The charged particle multiplicity for $N_{\rm{ch}}\geq 2$,
	\item[•] The mean charged transverse momentum vs. the
          multiplicity of charged
          particles for $p_{\perp}>500\,\rm{MeV}$ and $p_{\perp}>100\,\rm{MeV}$
	\item[•] The pion and the proton yield in the central rapidity region $|y|<0.5$.
\end{itemize}
The outcome of this tune is listed in Tab.\,\ref{table1} where we show the
parameter values that resulted in the lowest value of $\chi^2/N_{\rm dof}$ and
the values from the default tune of Herwig~7.1 without the baryonic colour
reconnection model.  The change in the colour reconnection algorithm and the
possibility to produce baryonic clusters results in an overall better
description of the considered observables.  While still being able to
accurately describe MB data we see the expected improvement in the charged
multiplicity distributions for the high multiplicity region which is due to
the baryonic colour reconnection.  The results of the tuning procedure will be
presented and discussed in the next section.

\section{Results}
\label{sec:results}
\begin{figure*}[t]
  \centering
  \includegraphics[width=0.49\textwidth]{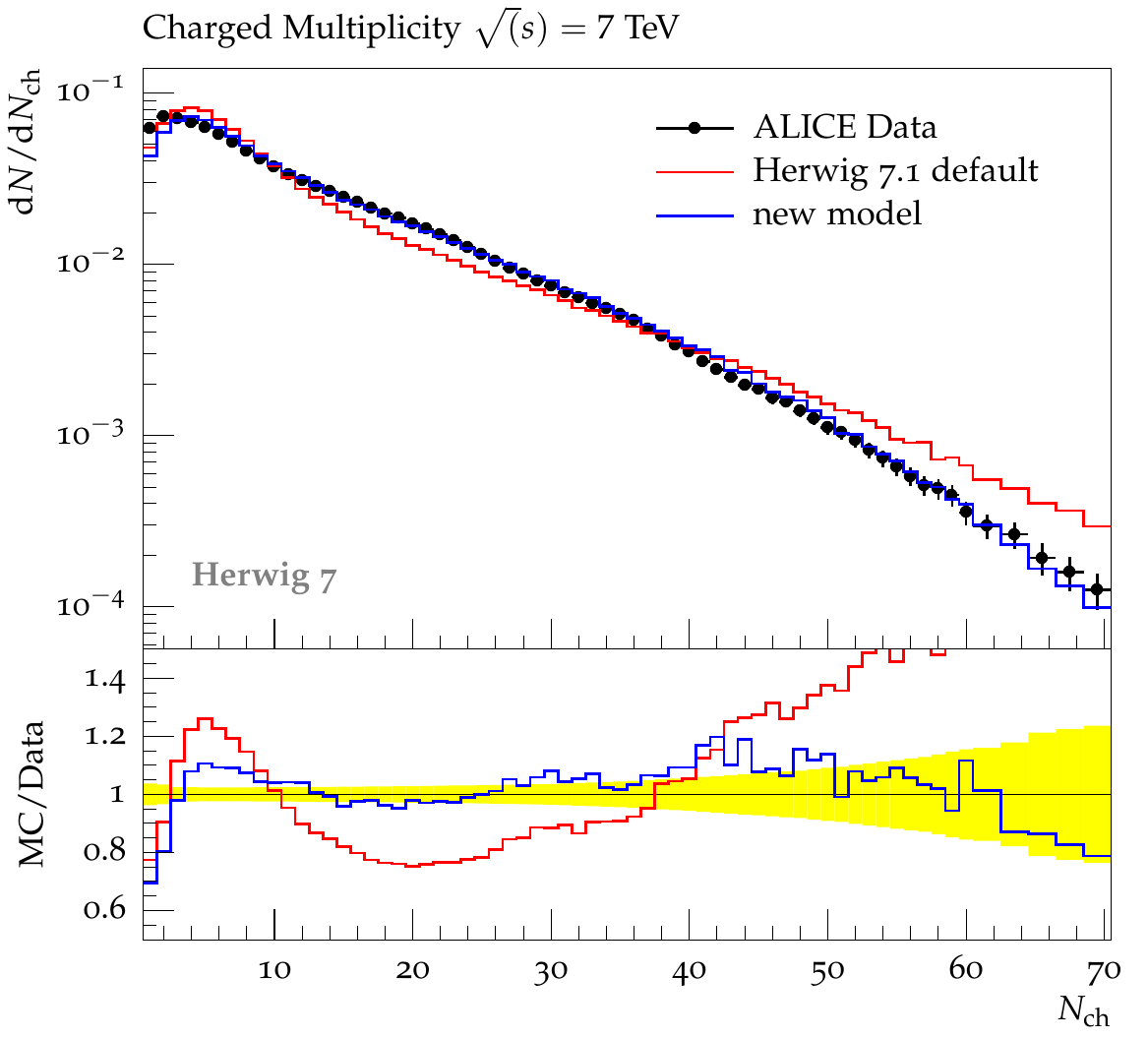}\hfill
  \includegraphics[width=0.49\textwidth]{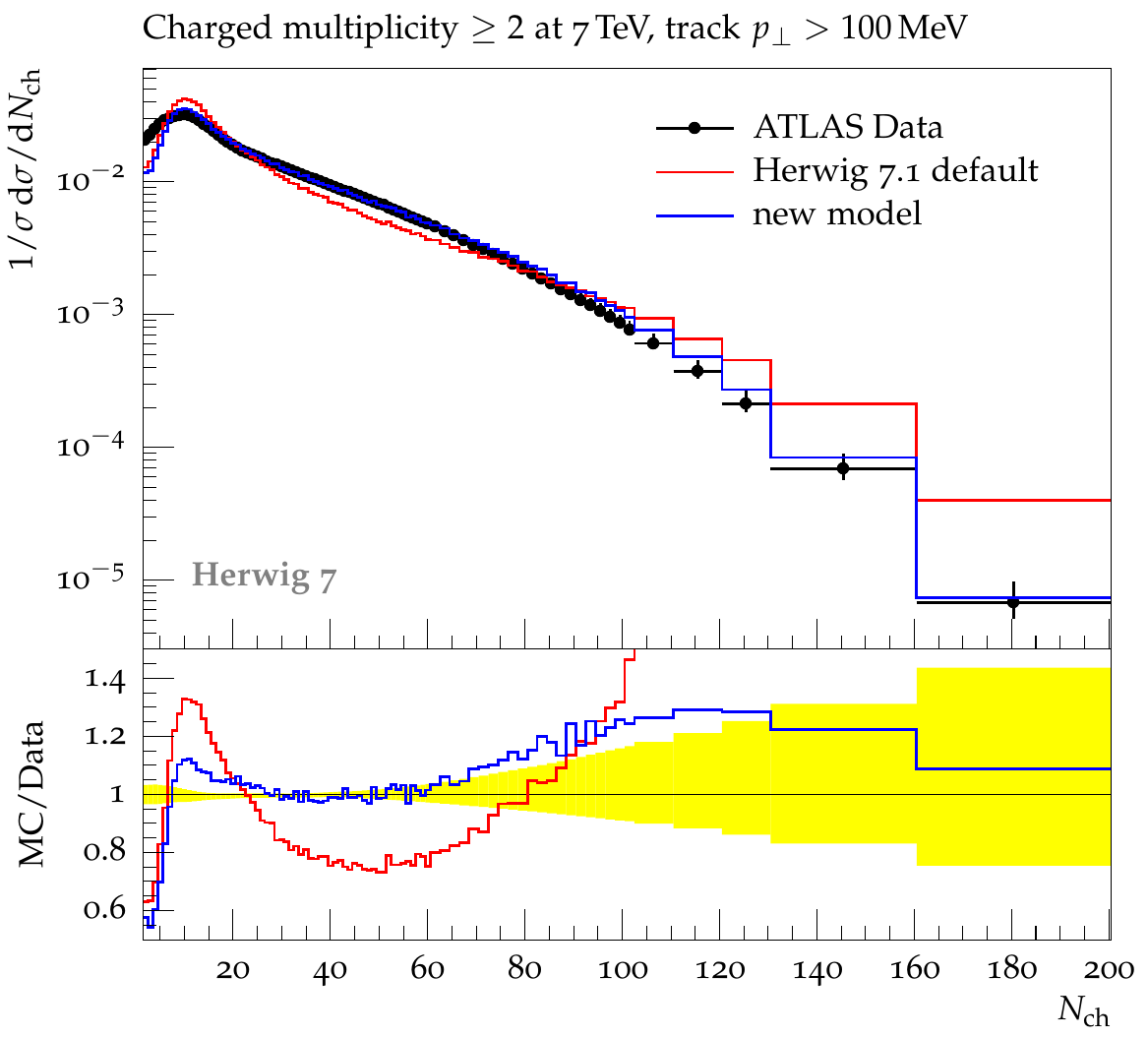}
  \caption{Multiplicity distributions as measured by ALICE for the central rapidity region
           $|\eta| < 1$ up to $N_{\mathrm{ch} = 70}$ \cite{Aamodt:2010pp}
           and ATLAS for $|\eta|<2.4$ up to $N_{\mathrm{ch}} = 200$ 
           for all particles with $p_{\perp} > 100 \,\mathrm{MeV}$\,\cite{Aad:2010ac}.
           We compare the old default colour reconnection model 
           with the new baryonic colour reconnection model.
    }
  \label{fig:results:nch}
\end{figure*}

Changes in the colour reconnection model are always \\deeply tied with the
peculiarities of the hadronization model. In principle one would have to
retune all parameters that govern hadronization in Herwig. This is usually
done in a very dedicated and long study with LEP data. We propose a simplified
procedure since little to no changes are expected with the extension to the
colour reconnection model in the $\rm{e}^++\rm{e}^-$ environment. At LEP the colour
structure of an event is not changed significantly through colour reconnection
since it is already well defined by the parton shower.  This was confirmed by
comparing the new model to a wide range of experimental data from LEP.  We
therefore keep the hadronization parameters that were tuned to LEP data (see
Refs.\,\cite{Bahr:2008pv,Bellm:2017bvx}) at their default values.
We also note that this does not replace a dedicated study concerned 
with the tuning and validation of hadronization parameters. Especially at
pp collisions a different model for colour reconnection leads to changes in
the interplay between the clusters and the hadronization in an
unforeseeable way. A possible way out of this dilemma would be to make a
distinguished LHC tune and compare the results with LEP.
Nonetheless we restrain ourselves to the explained simplified method
in order to make qualitative statements about the new model for colour reconnection.
The new model with the tuned parameters improves the description of all
observables considered in the tuning procedure.
The effect of the baryonic colour reconnection was already demonstrated in
Fig.\,\ref{fig:nch}.  In Fig.\,\ref{fig:results:nch} we show the same
distribution of the charged-particle multiplicity for the central region $|y|
< 1$ with the tuned parameter values.  Again we see the expected fall off for
high multiplicities.  The new model is able to describe the whole region
fairly well compared to the old model.  Only the low multiplicity region
$n<10$ is overestimated by a factor of $\approx 10\%$ and for $n<5$ underestimated. 
In Fig.\,\ref{fig:results:nch} we also show a similar observable
for a wider rapidity region $|y|<2.4$ and up to $n=200$ as measured by
CMS\,\cite{Aad:2010ac}.  Again the central multiplicity region shows a
significant improvement.  For multiplicities $n>80$ we note a slight
overestimation of the data but are still within error bars.
 
This can be understood quite simply: the more activity in an event, the more
likely it becomes that a cluster configuration that leads to baryonic
reconnection is found.  The high multiplicity events therefore exhibit a
disproportionately large fraction of baryonic reconnections. Due to the 
highly restricted phase space for the production of baryons from baryonic clusters
less particles are produced than with mesonic clusters of the same invariant mass, 
which lowers the charged multiplicity.

We also observe the proposed change in mesonic and baryonic activity in the
$p_{\perp}$ spectra of pions and protons.  Especially the $\rm{p}/\pi$ ratio
and the $p_{\perp}$ distributions improve significantly which should be
considered first in a model that tries to explain flavour multiplicities.
When looking at the $p_{\perp}$ distributions of $\rm{K}$ and $\rm{\Lambda}$ we see
that none of the performed tunes is able to capture the essence of these
distributions correctly which is no surprise since we have not touched or
altered the production mechanism of strange particles.  We merely observe a
small increase in the $p_{\perp}$ distribution of $\rm{\Lambda}$ baryons due to the
baryonic reconnection.  Changes that affect the hadronization model usually
have severe consequences for the hadronization parameters.  We restrict
ourselves to the parameters that are responsible for strangeness production and
allow one additional source of strangeness in the event generator work flow.
We exploit the freedom one is given by LEP observables for the probability to
select a strange quark during cluster fission \mbox{\tt PwtSquark} and
additionally allow non-perturbative gluon splitting into strange quarks with a
given probability \mbox{\tt SplitPwtSquark}.  In a second tuning procedure we
consider these two additional parameters and also tune to the $p_{\perp}$
distribution of the $\pi^+\pi^-$, $\rm{K}^++\rm{K}^-$, $\rm{p}+\bar{\rm{p}}$
yields in inelastic pp collisions at $7\,\mathrm{TeV}$\,\cite{Adam:2015qaa}
and the $p_{\perp}$ distribution of $\rm{\Lambda}$\,\cite{Khachatryan:2011tm}.  The
parameter values that were obtained in the tuning are listed in
Tab.~\ref{table2}.

In addition to the tuned observables, many hadron flavour observables 
which were not considered in the tuning procedure show a significant improvement as well.
In order to compare the different effects from the new colour reconnection
model and the possibility to produce strange quarks during gluon splitting we
made runs with the default model (Herwig 7.1 default), the pure baryonic
colour reconnection model (baryonic reconnection), one run where we allow the
gluons to split into strange quarks ($g\to s\bar{s}$ splittings) and use the
old colour reconnection model and a run where we use both extensions and the
parameters that we obtained from the tuning (new model).

\begin{figure*}[t]
  \centering
  \includegraphics[width=0.49\textwidth]{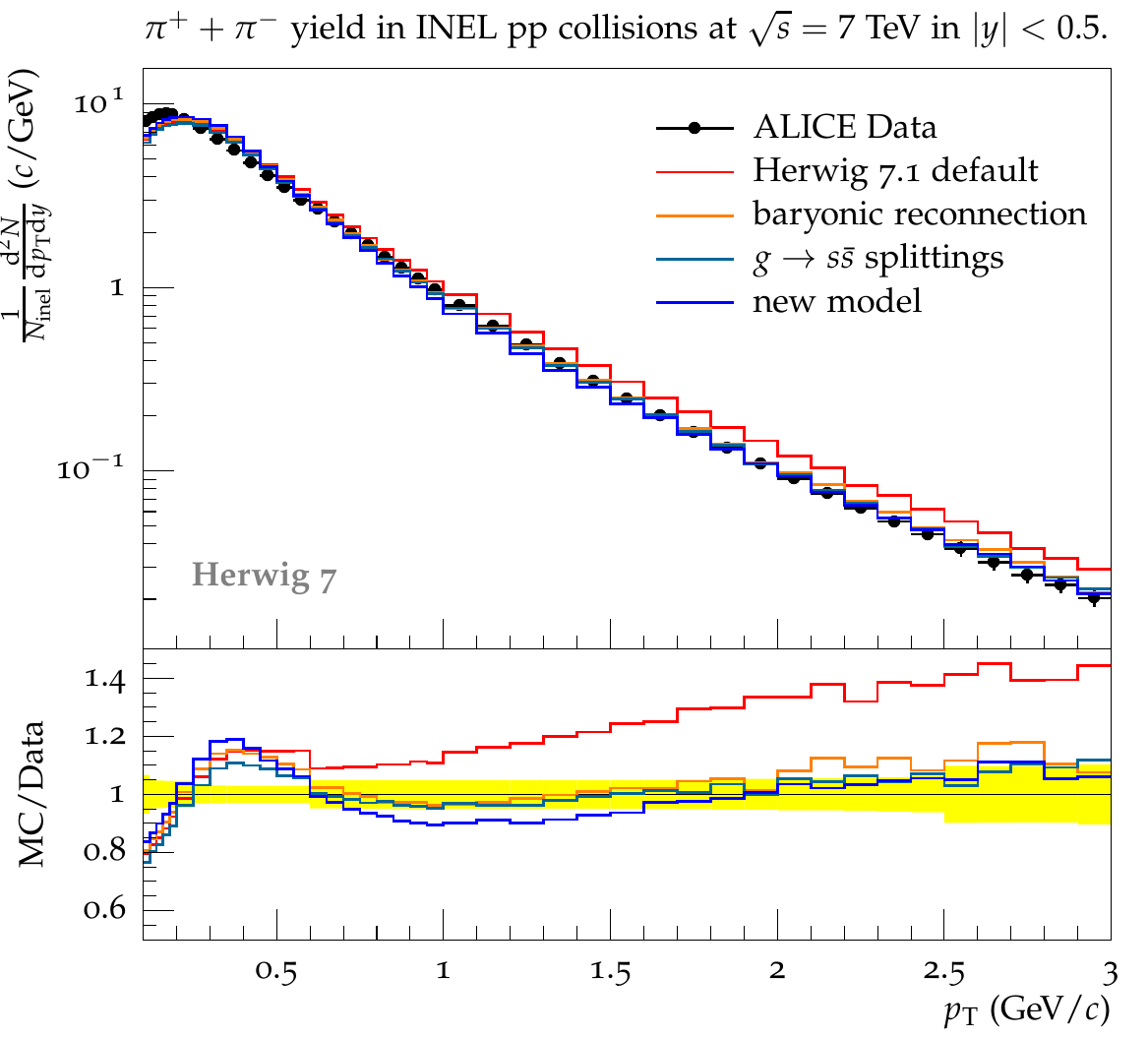}\hfill
  \includegraphics[width=0.49\textwidth]{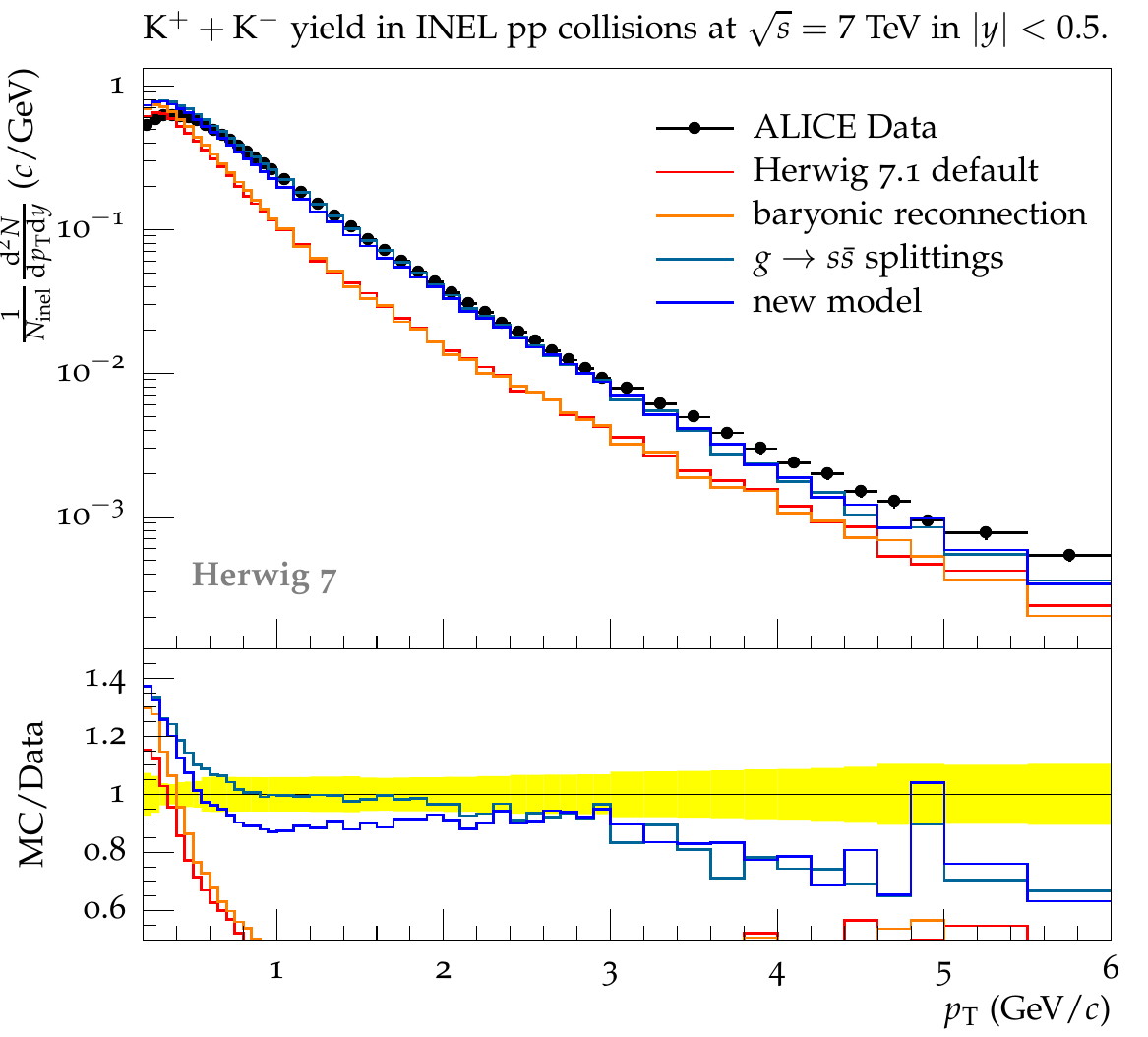}
  \caption{ The transverse momentum spectra for $\mathrm{\pi^++\pi^-}$ and $\mathrm{p+\bar{p}}$ as measured by 
            ALICE at $\sqrt{s}=7\,\mathrm{TeV}$ \cite{Adam:2015qaa} in the very central rapidity region $|y|<0.5$.
    }
  \label{fig:results:yields1}
\end{figure*}

\begin{figure}[t]
\centering
\includegraphics[width=9cm]{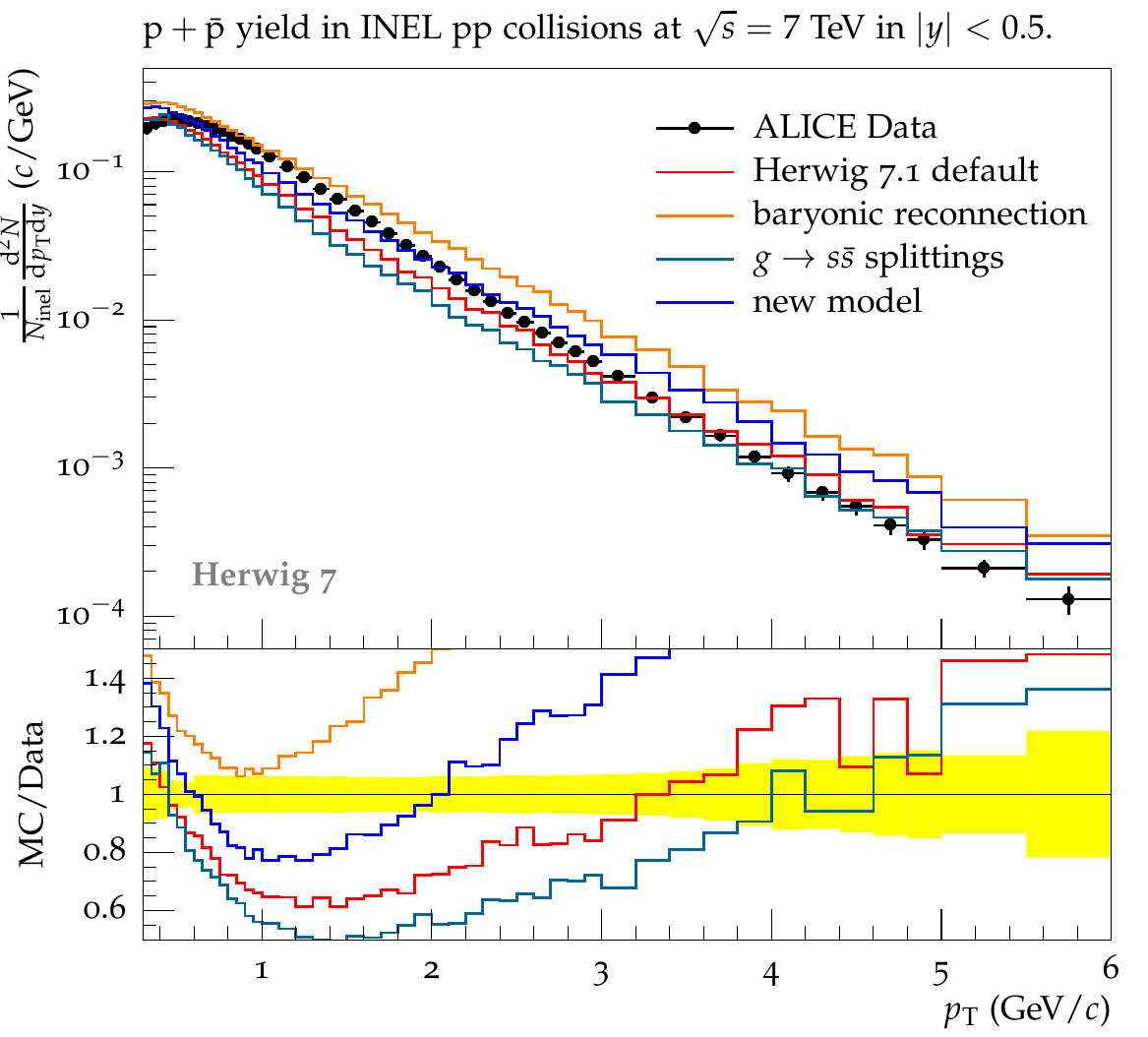}
\caption{ The transverse momentum spectrum for $\mathrm{p+\bar{p}}$ as
  measured by ALICE at $\sqrt{s}=7\,\mathrm{TeV}$ \cite{Adam:2015qaa} in the
  very central rapidity region $|y|<0.5$.  }
\label{fig:results:baryonyield}
\end{figure}

\begin{figure*}[th]
  \centering
  \includegraphics[width=0.49\textwidth]{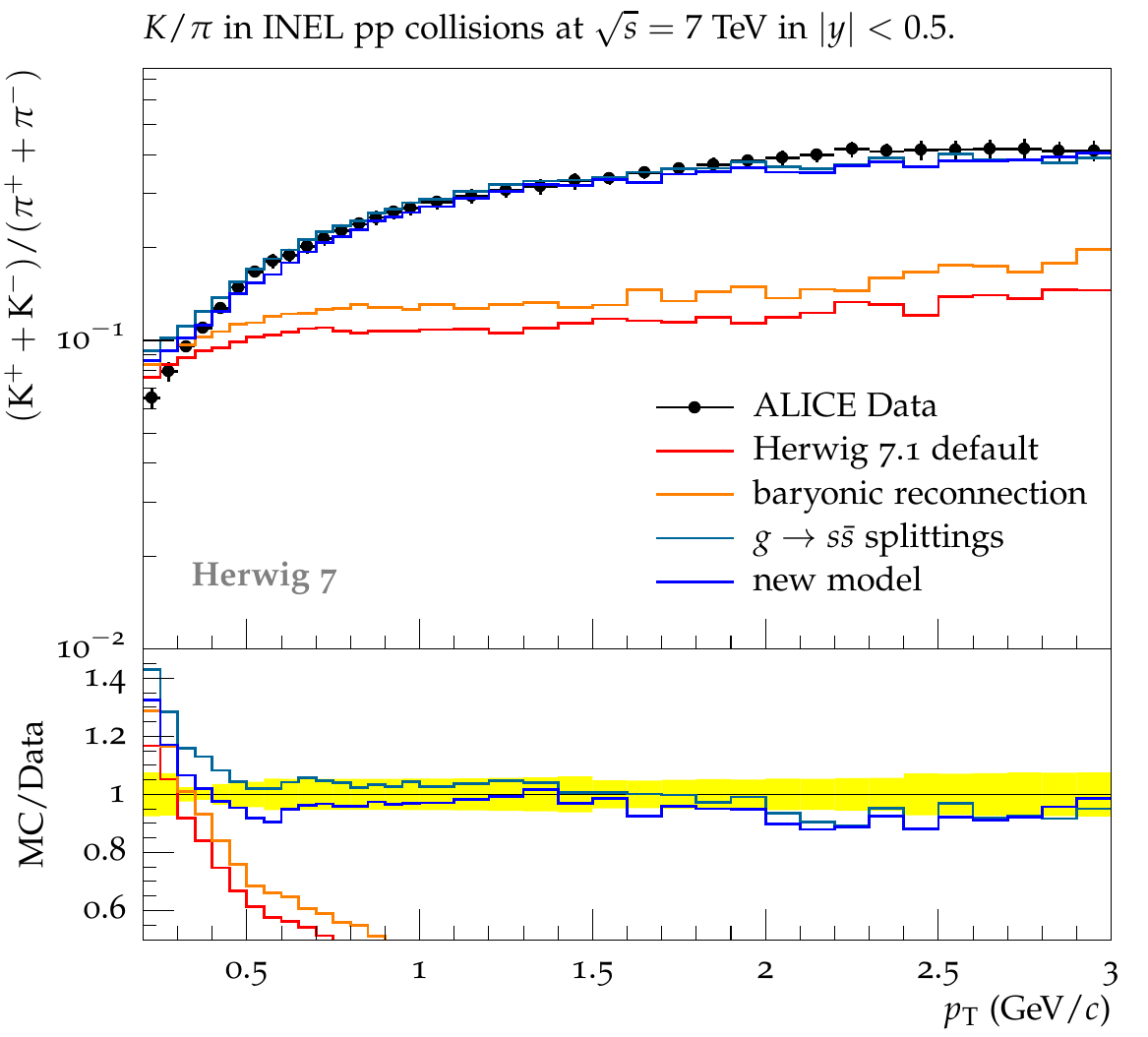}\hfill
  \includegraphics[width=0.49\textwidth]{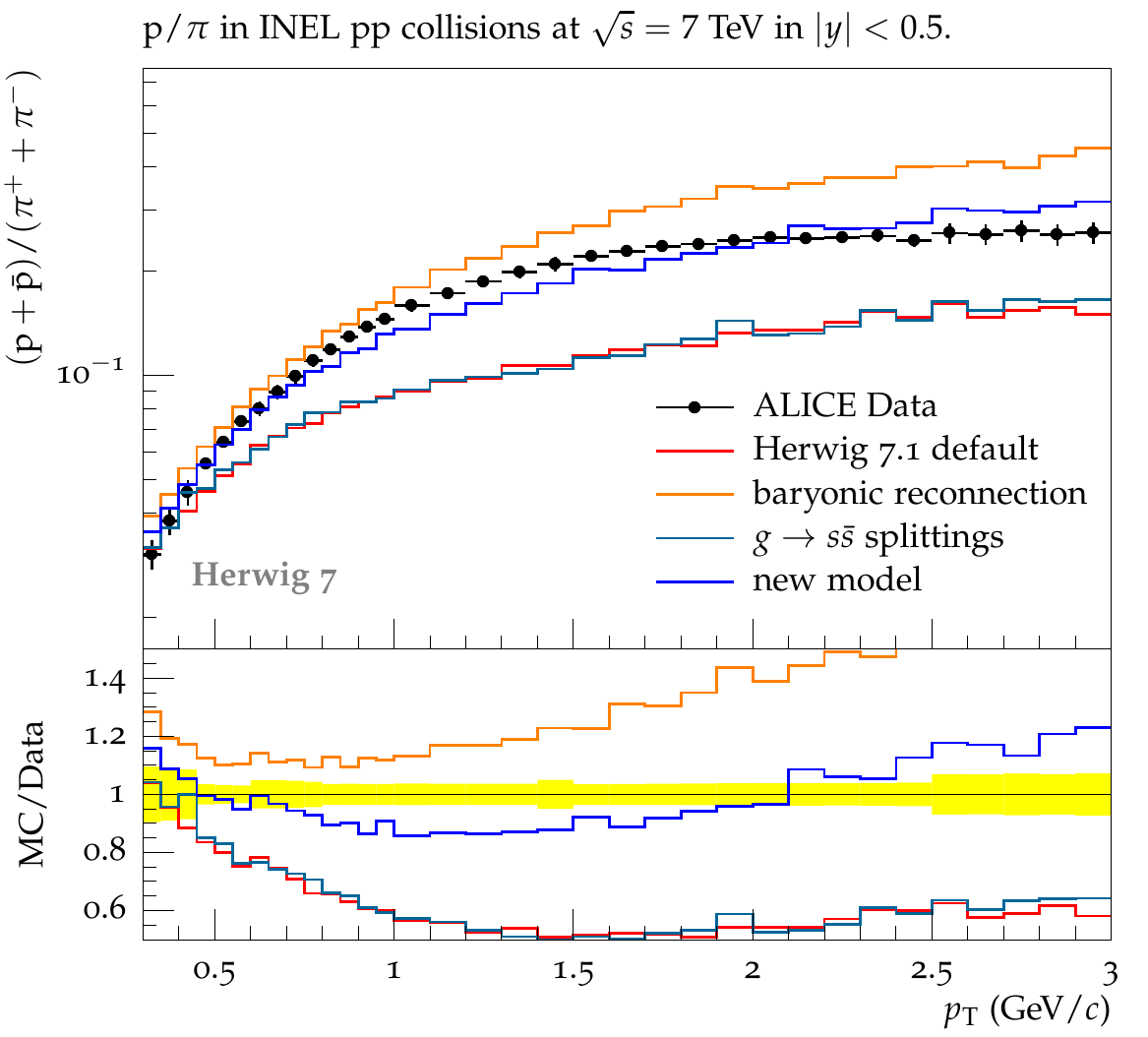}
  \caption{Transverse momentum spectra for the ratios $\mathrm{p/\pi}$ and
    $\mathrm{K/\pi}$ as measured by ALICE at $\sqrt{s}=7\,\mathrm{TeV}$
    \cite{Adam:2015qaa} in the very central rapidity region $|y|<0.5$.  }
  \label{fig:results:ratios}
\end{figure*}

In Fig.\,\ref{fig:results:yields1} we show the $p_{\perp}$ distributions of
$\pi$ and $\rm{K}$ in the central rapidity region as measured by
ALICE\,\cite{Adam:2015qaa} and in Fig.\,\ref{fig:results:baryonyield} the
corresponding $\rm{p+\bar{p}}$ distribution.  While all options improve the
description of pions we see that the $\rm{K}$ distribution can only be
described if we take the additional source of strangeness into account.  The
proton $p_{\perp}$ distribution is mainly driven by baryonic reconnection.
The rate increases for all $p_{\perp}$ regions but we overshoot the data by a
large factor for $p_{\perp}>3\,\mathrm{GeV}$ and for the very low $p_{\perp}$
region.  Since all options show the same trend this might indicate some
problems with the hard part of the MPI model which dominates
$p_{\perp}>3\,\mathrm{GeV}$.  In Fig.\,\ref{fig:results:ratios} we consider
the hadron ratios $\mathrm{K}/\pi$ and $\mathrm{p}/\pi$.  The new model does a
significant better job in describing the data and only the combined effect of
the enhanced baryon production through the change in the colour reconnection
model and gluon splitting into strange quarks is able to give a satisfying
description of both observables.

\begin{figure*}[th]
  \centering
  \includegraphics[width=0.49\textwidth]{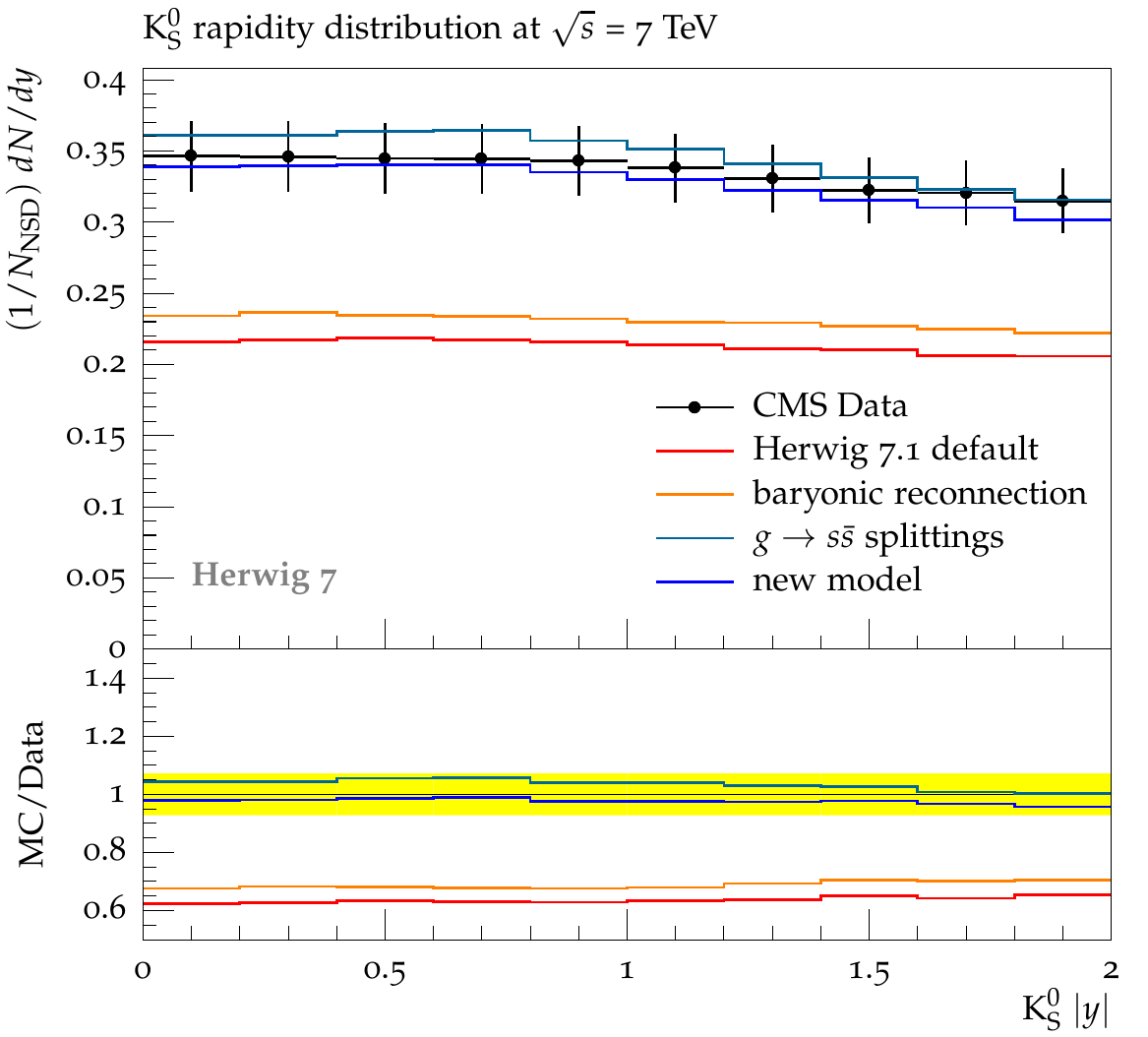}\hfill
  \includegraphics[width=0.49\textwidth]{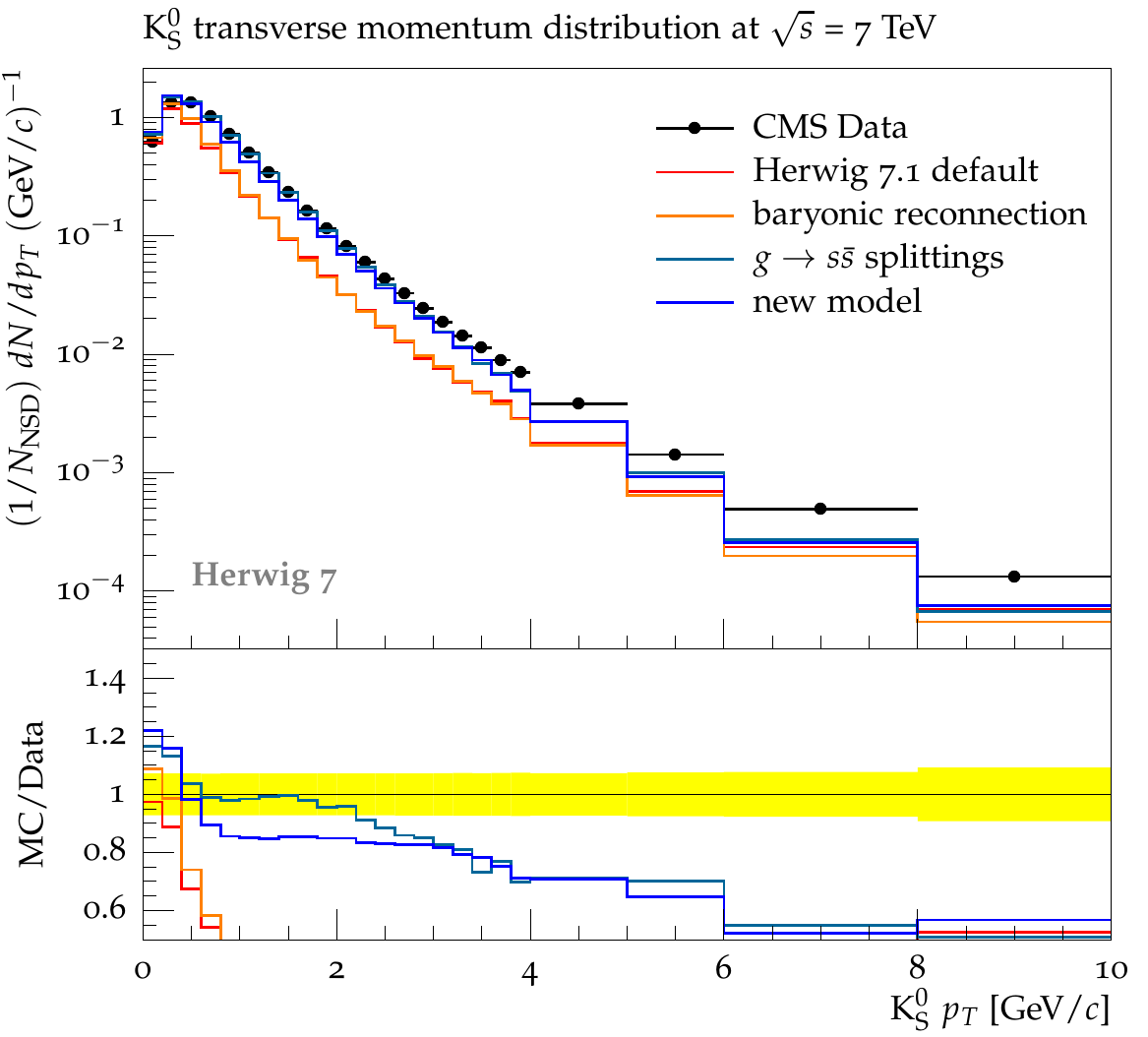}
  \caption{The $\mathrm{K_S^0}$ rapidity and $p_{\perp}$ distribution as measured by CMS at $\sqrt{s}=7\,\mathrm{TeV}$\,\cite{Khachatryan:2011tm}.
    }
  \label{fig:results:kaons}
\end{figure*}

\begin{figure*}[th]
  \centering
  \includegraphics[width=0.49\textwidth]{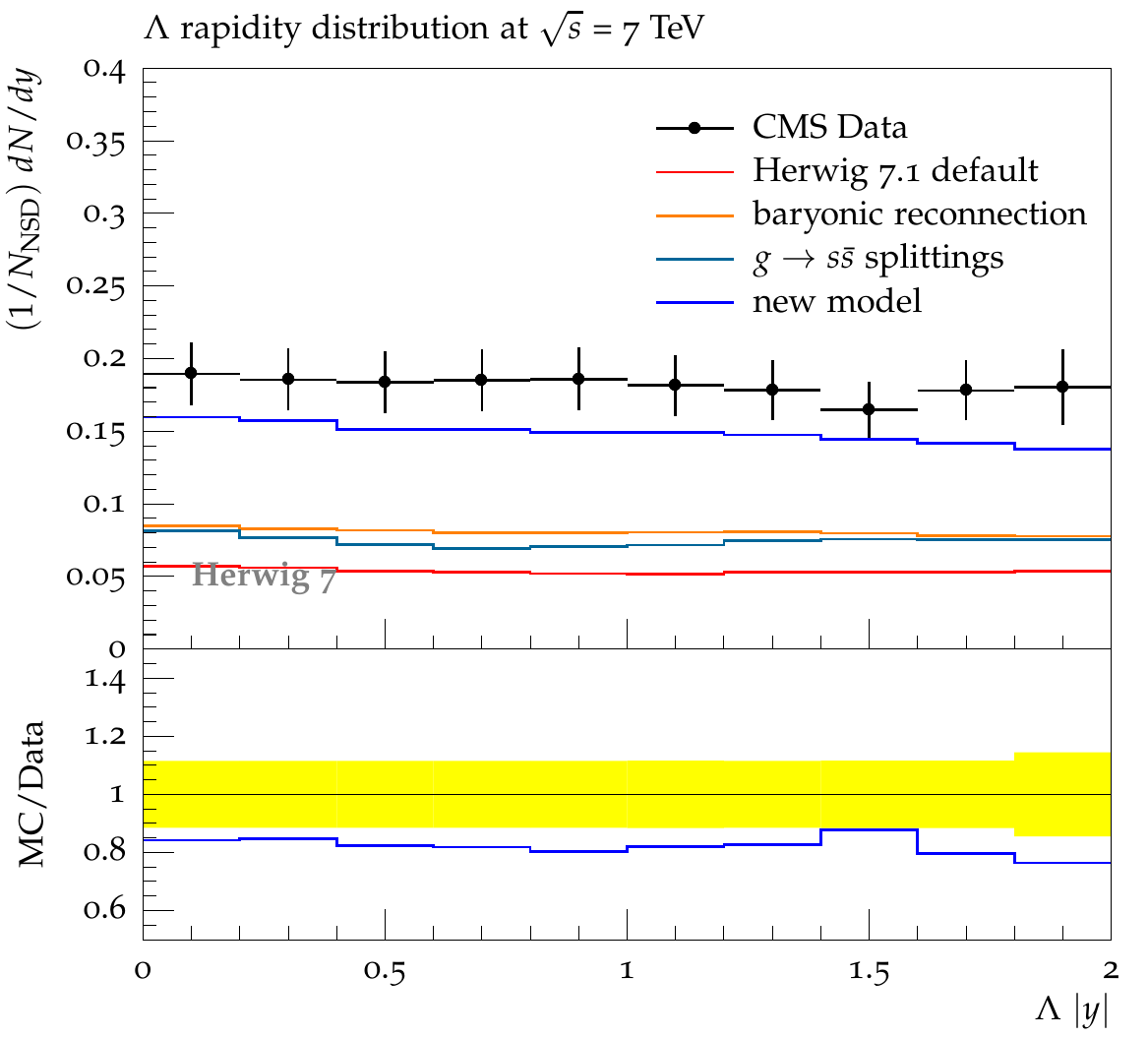}\hfill
  \includegraphics[width=0.49\textwidth]{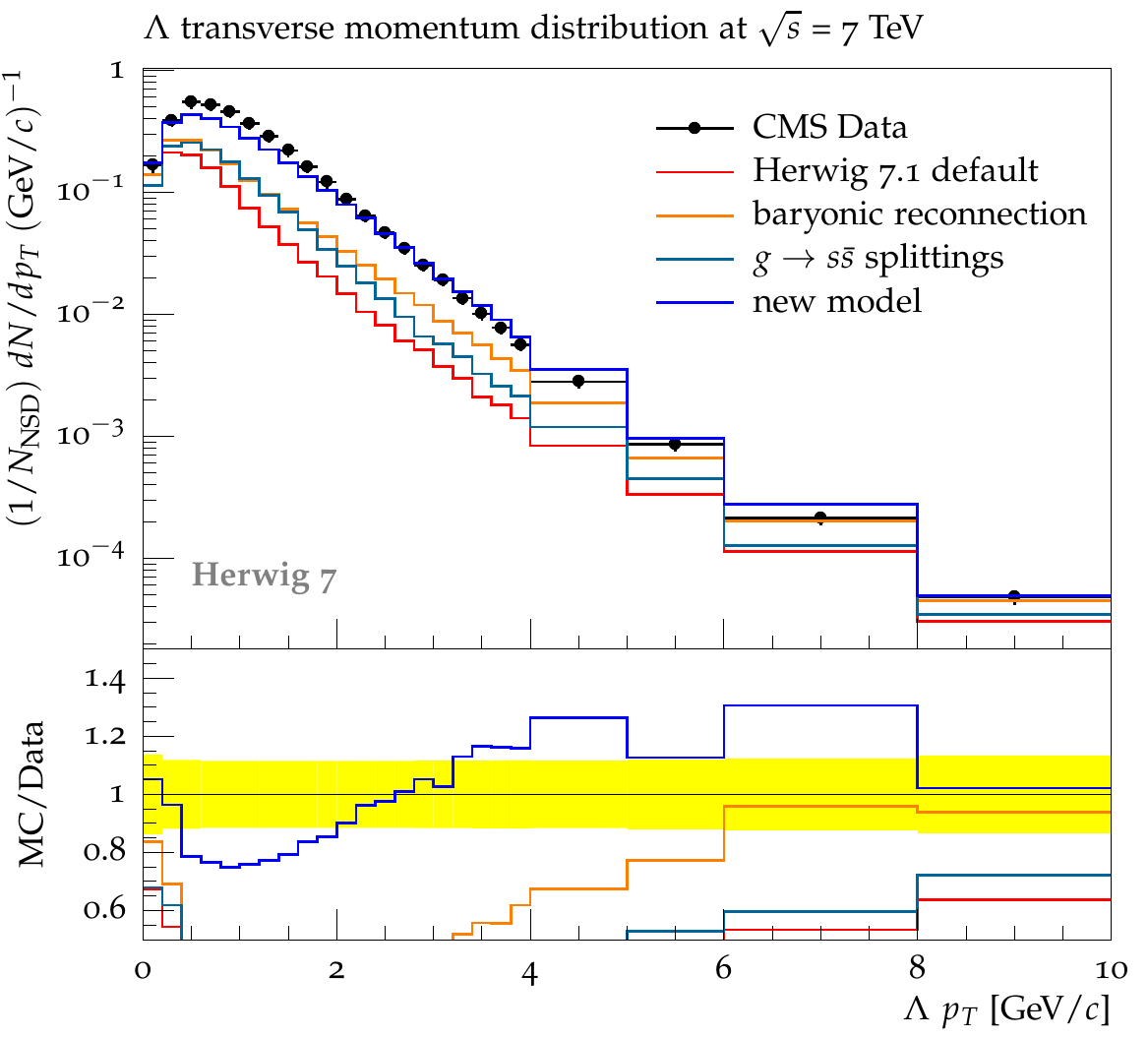}
  \caption{The $\mathrm{\Lambda}$ rapidity and $p_{\perp}$ distribution as measured by CMS at $\sqrt{s}=7\,\mathrm{TeV}$\,\cite{Khachatryan:2011tm}.
    }
  \label{fig:results:lamdas}
\end{figure*}

\begin{figure*}[th]
\centering
\includegraphics[width=9cm]{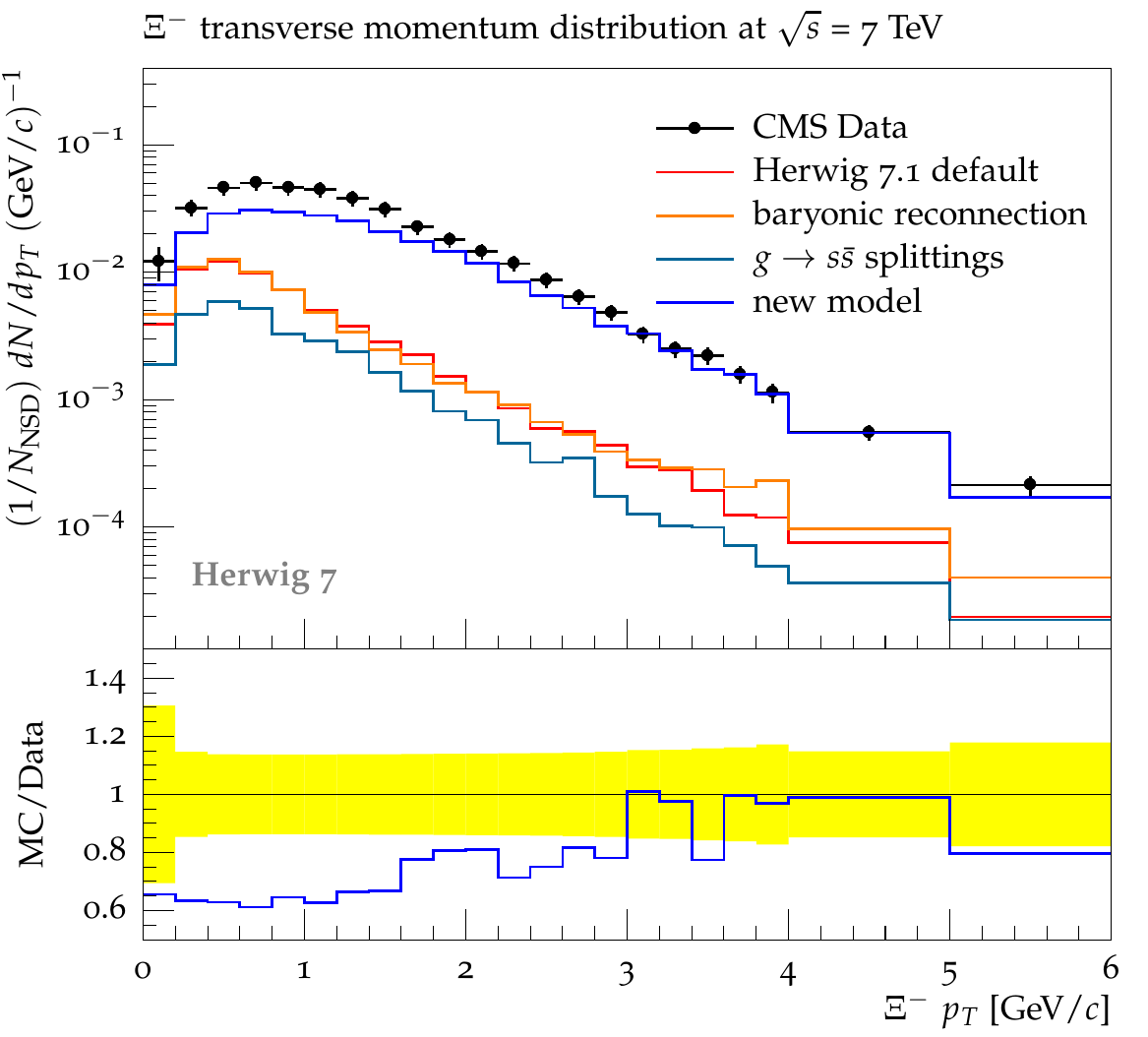}
\caption{The $\mathrm{\Xi^-}$ $p_{\perp}$ distribution as measured by CMS at $\sqrt{s}=7\,\mathrm{TeV}$\,\cite{Khachatryan:2011tm}.
}
\label{fig:results:xis}
\end{figure*}

\begin{figure*}[th]
  \centering
  \includegraphics[width=0.49\textwidth]{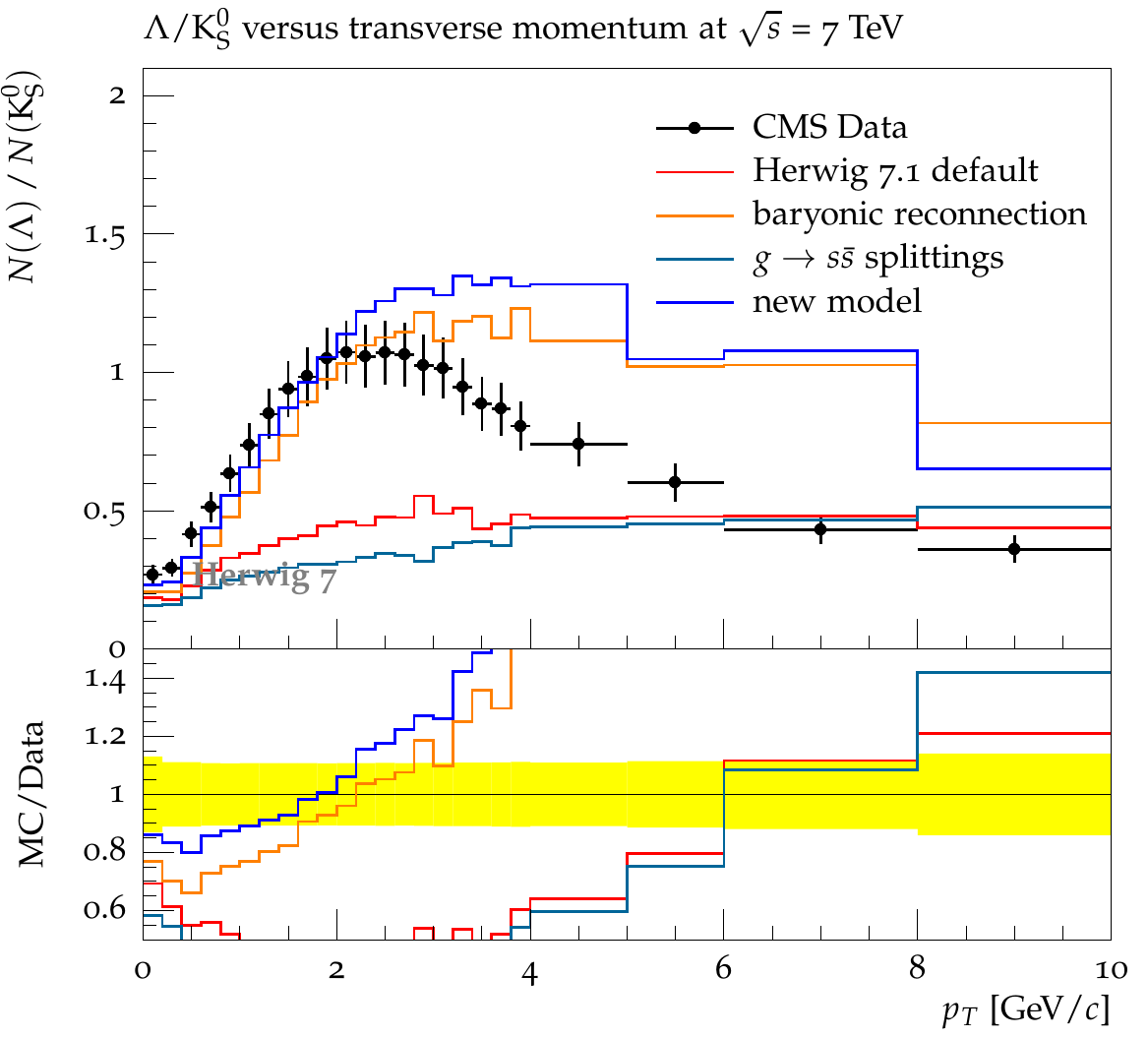}\hfill
  \includegraphics[width=0.49\textwidth]{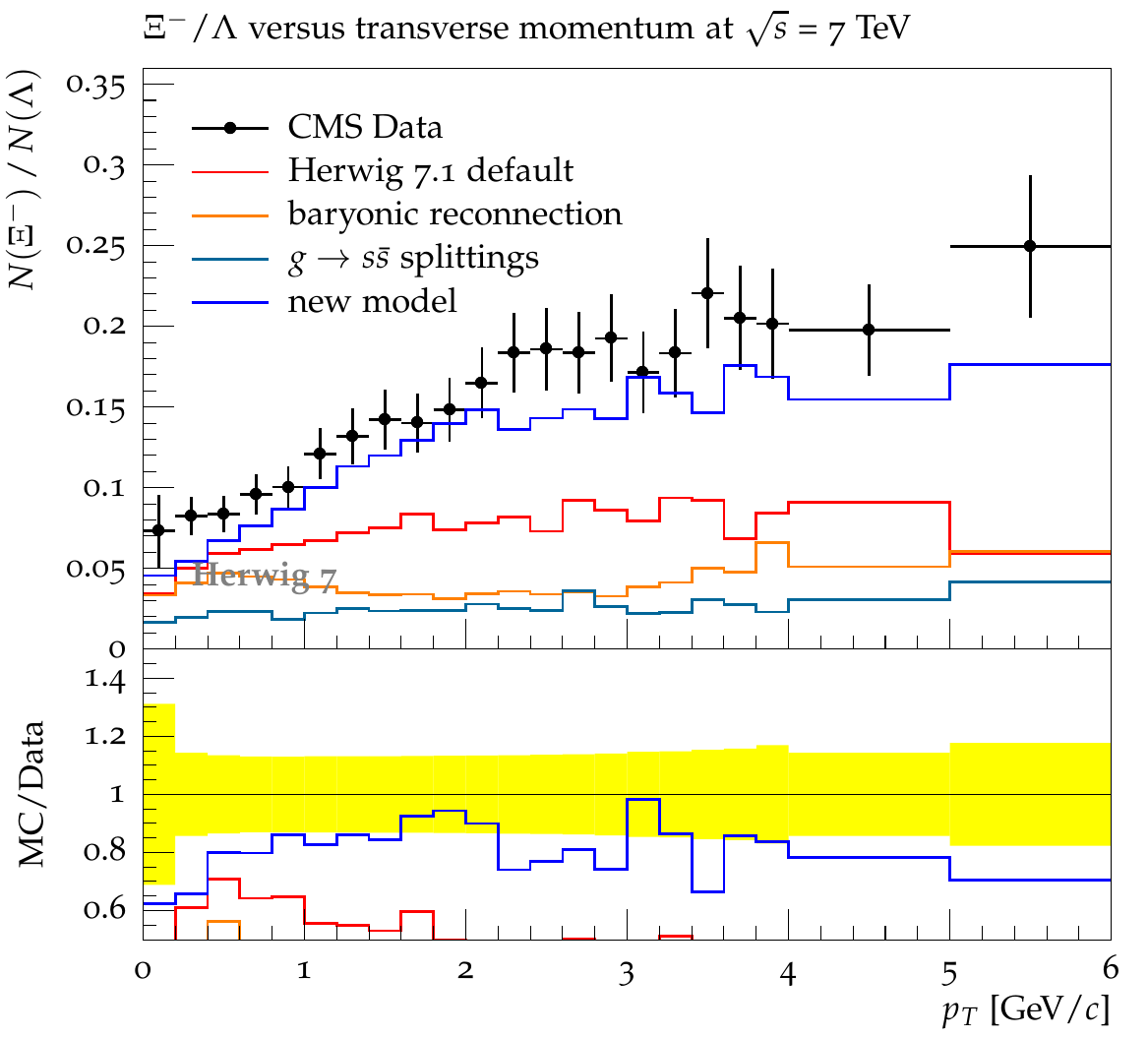}
  \caption{ The $\mathrm{\Lambda/K_S^0}$ and the $\mathrm{\Xi^-/\Lambda}$ $p_{\perp}$ distribution as measured by CMS at $\sqrt{s}=7\,\mathrm{TeV}$ \cite{Khachatryan:2011tm}.
    }
  \label{fig:results:ratios-strangeness}
\end{figure*}

In Figs.\, \ref{fig:results:kaons}, \ref{fig:results:lamdas},
\ref{fig:results:xis}, \ref{fig:results:ratios-strangeness} we compare the
model to $\sqrt{s} = 7\, \mathrm{TeV}$ data from
CMS\,\cite{Khachatryan:2011tm} for the strange flavour observables of
$\mathrm{K_S^0}$, $\mathrm{\Lambda}$ and $\mathrm{\Xi^-}$.  The new model
improves the description for all observables published in this analysis.
Again we show the effects of the different contributions and note that the
best description can only be achieved by a combination of baryonic colour
reconnection and gluon splitting into strange quarks (new-tune).  The
$\mathrm{\Lambda/K_S^0}$ distribution shows a good description in the turn on
region but the high $p_{\perp}$ tail is not well described.  A similar
observation was made with Pythia in\,\cite{Christiansen:2015yqa}.
Surprisingly the $\mathrm{\Xi^-/\Lambda}$ distribution is able to capture the
general trend but due to large errors in the high $p_{\perp}$ region it is
difficult to draw conclusions.  We see significant improvement in the
description of hadron flavour observables.  Especially the rapidity
distributions and the particle ratios $\mathrm{\Lambda/K_S^0}$ and
$\mathrm{\Xi^-/\Lambda}$ show a large enhancement compared to the default
model.  Again we point out the interplay between baryonic colour reconnection
and the strangeness production mechanism which is responsible for the
improvement in the description of the heavy baryons $\mathrm{\Lambda}$ and
$\mathrm{\Xi^-}$.

In a recent analysis by ALICE a significant enhancement of strange to non-
strange hadron production with increasing particle multiplicity in pp
collisions was observed\,\cite{ALICE:2017jyt}.  Since we are developing a
model that incorporates strangeness production and the enhanced production of
baryons it is instructive to compare our model to the data published by the
ALICE collaboration.

\begin{figure*}[th]
  \centering
  \includegraphics[width=0.49\textwidth]{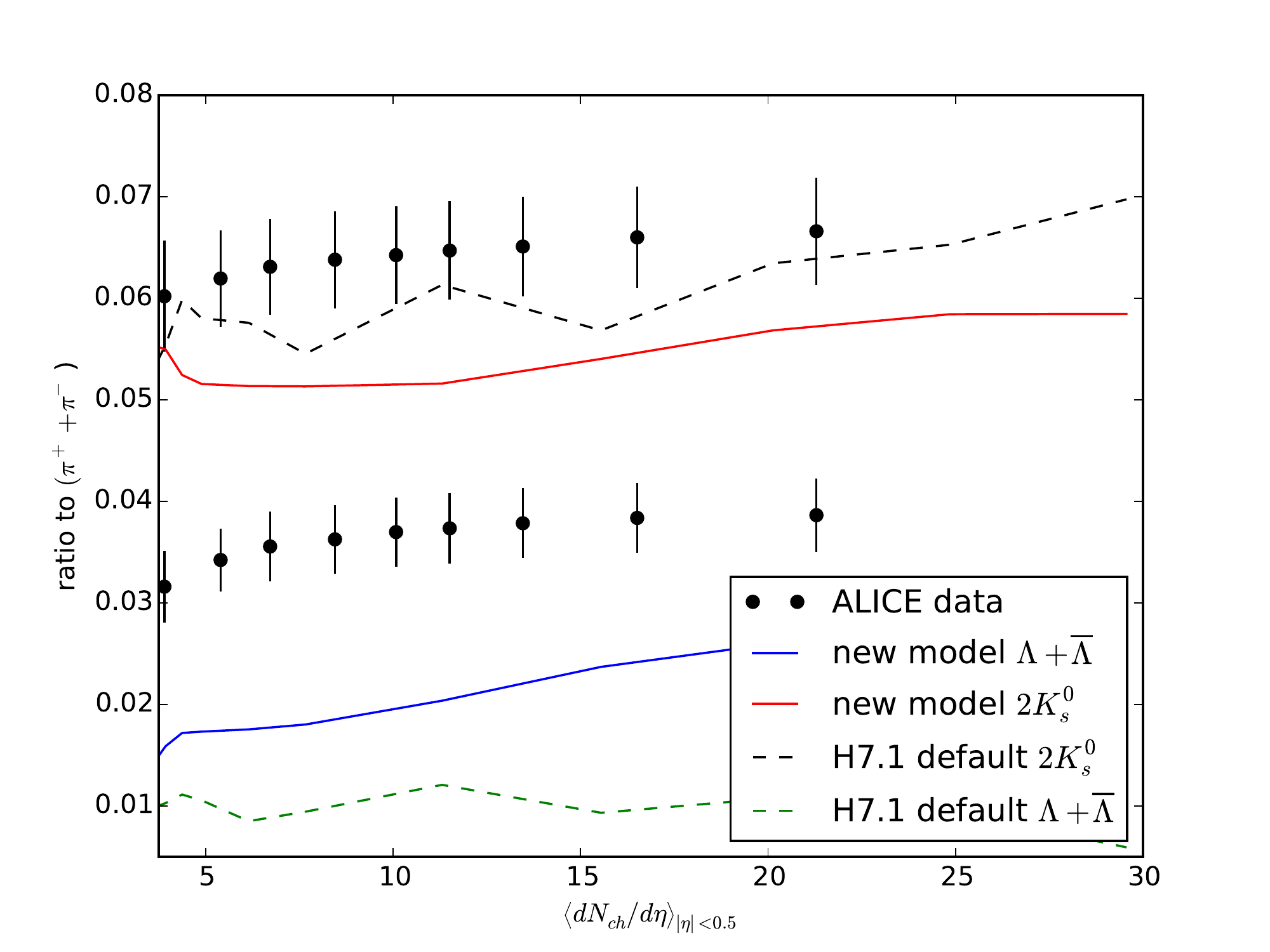}\hfill
  \includegraphics[width=0.49\textwidth]{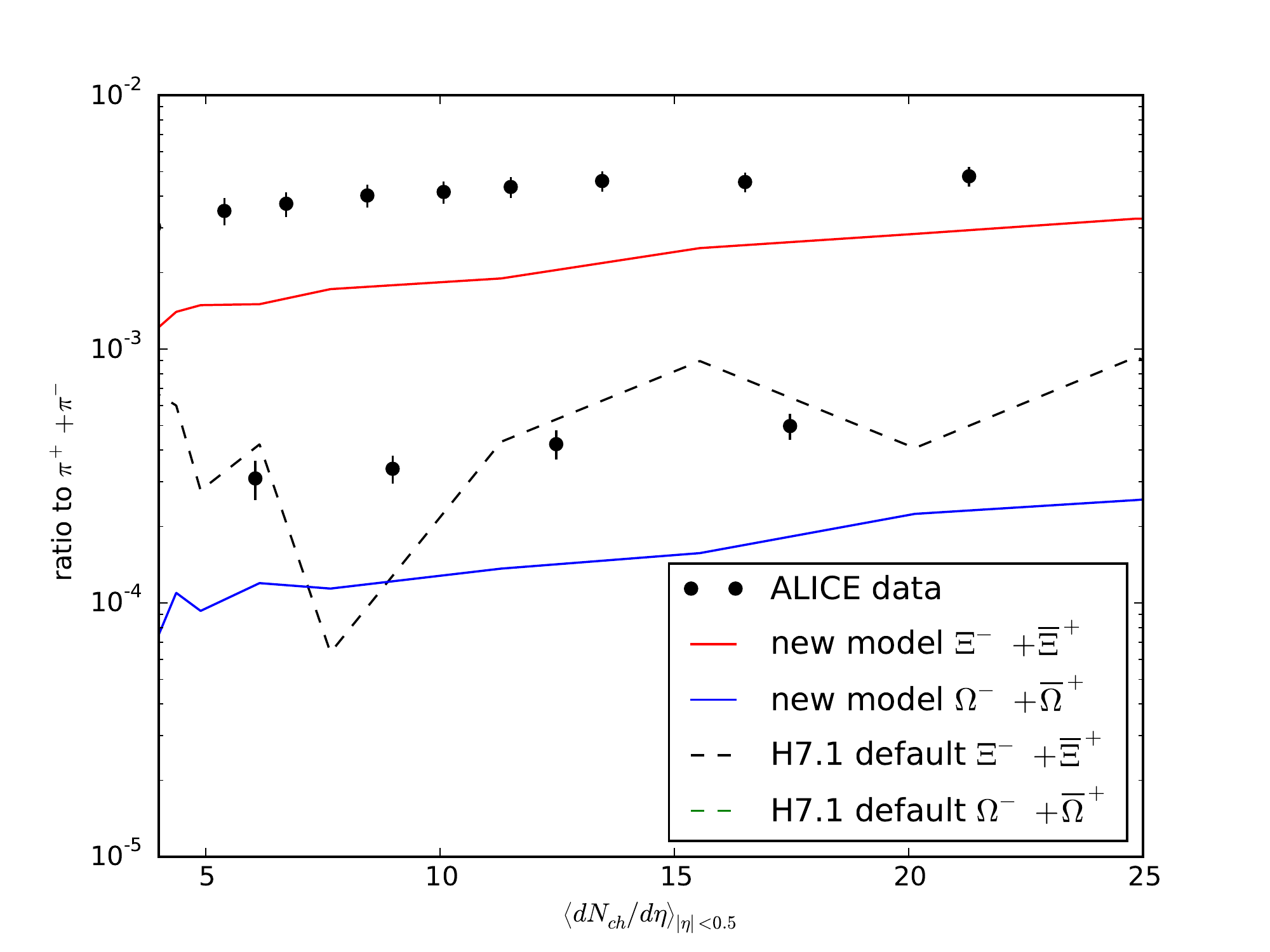}
  \caption{Integrated (multi)-strange particle yield ratios to $\mathrm{\pi^+}+\mathrm{\pi^-}$ 
as a function of $\langle \mathrm{d}N_{ch}/\mathrm{d}\eta \rangle$ for $|\eta|<0.5$. 
The values from the ALICE analysis\,\cite{ALICE:2017jyt} are compared to calculations from Herwig 7.1 with the 
old model and the new model for colour reconnection.}
  \label{fig:results:alice1}
\end{figure*}

In Fig.\,\ref{fig:results:alice1} we show we show the
$(\mathrm{\Lambda}+\mathrm{\bar{\Lambda}})/(\mathrm{\pi^+}+\mathrm{\pi^-})$
and the $\mathrm{K_s^0}/(\mathrm{\pi^+}+\mathrm{\pi^-})$ ratio for the old
model and the new model for colour reconnection. While the
$\mathrm{K_s^0}/(\mathrm{\pi^+}+\mathrm{\pi^-})$ ratio is reasonably well
described, the $\rm{\Lambda/\pi}$ ratio is underestimated by both, the old and
the new model. The new model, on the other side is able to capture the general
trend of the observable and describe the rise of the fraction of
$\rm{\Lambda}$ baryons with respect to increasing particle multiplicity
correctly. Also the increase in the fraction of multi-strange baryons
$(\mathrm{\Xi^-}+\mathrm{\bar{\Xi}^+})$ and
$(\mathrm{\Omega^-}+\mathrm{\bar{\Omega}^+})$ can qualitatively be described
by the new model as is also shown in Fig.\,\ref{fig:results:alice1}.  Note
that the Herwig 7.1 default model did not produce clusters heavy enough in
order to account for the production of any $\mathrm{\Omega}$-Baryons.

\begin{figure*}[th]
  \centering
  \includegraphics[width=0.49\textwidth]{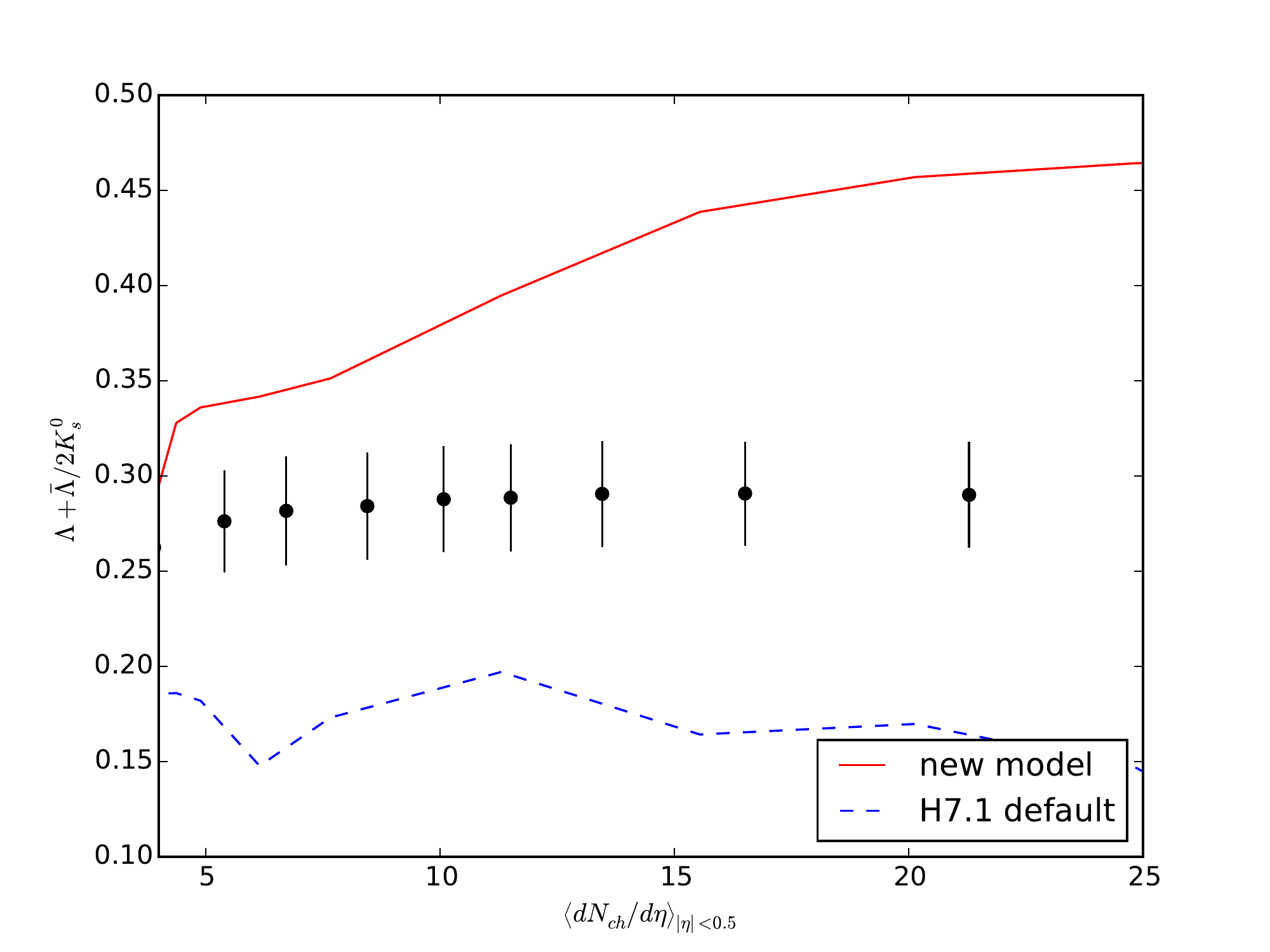}\hfill
  \includegraphics[width=0.49\textwidth]{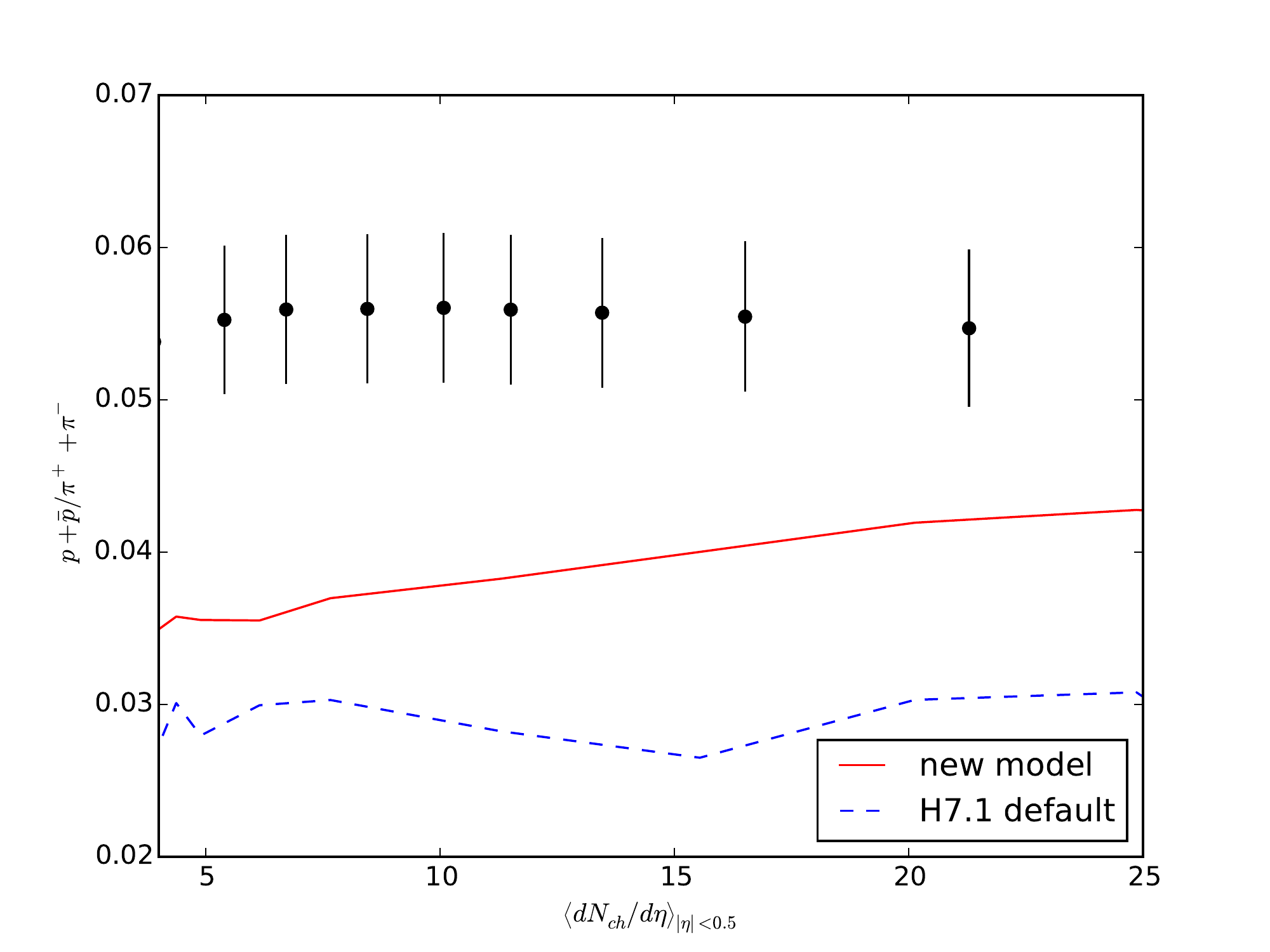}
  \caption{Integrated particle yield ratio of $\mathrm{\Lambda}+\mathrm{\bar{\Lambda}}/2\mathrm{K_s^0}$ 
and $\mathrm{p}+\bar{\mathrm{p}}/\pi^++\pi^-$ as a function of $\langle \mathrm{d}N_{ch}/\mathrm{d}\eta \rangle$ 
for $|\eta|<0.5$. The values from the ALICE analysis\,\cite{ALICE:2017jyt} are compared to calculations from Herwig 7.1 
with the old model and the new model for colour reconnection.}
  \label{fig:results:alice2}
\end{figure*}

In Fig.\,\ref{fig:results:alice2} we show the ratio of the single strange
particles $(\mathrm{\Lambda}+\mathrm{\bar{\Lambda}})/2\mathrm{K_s^0}$ and the
ratio of $\mathrm{p}+\mathrm{\bar{p}}/(\mathrm{\pi^+}+\mathrm{\pi^-})$ which do
not contain any strange quarks.  In both cases Herwig is not able to describe
the data correctly.  We see an disproportional large increase of the
$(\mathrm{\Lambda}+\mathrm{\bar{\Lambda}})/2\mathrm{K_s^0}$ ratio with
increasing multiplicity which is due to the nature of our colour reconnection
model, that enhances the production of baryons with increasing event
multiplicity.  This increase is not as pronounced in the
$\mathrm{p}+\mathrm{\bar{p}}/(\mathrm{\pi^+}+\mathrm{\pi^-})$ ratio but still
is present.

The comparison with ALICE data gives us interesting insights into our new
model for colour reconnection. Although we manage to capture the general trend
in the increasing fraction of (multi)-strange baryons to non-strange mesons,
we are not able to reproduce all aspects of the observations made by
ALICE. Especially the $\rm{\Lambda/K}$ ratio and the $\rm{p/\pi}$ ratios are
not well described since we have a mechanism that increases the production of
baryons in total and does not lead to an increase in the strange baryons only.
 
\subsection{Spectra of cluster masses}
\label{subsec:clustermasses}

\begin{figure}[t]
\centering
\includegraphics[width=8cm]{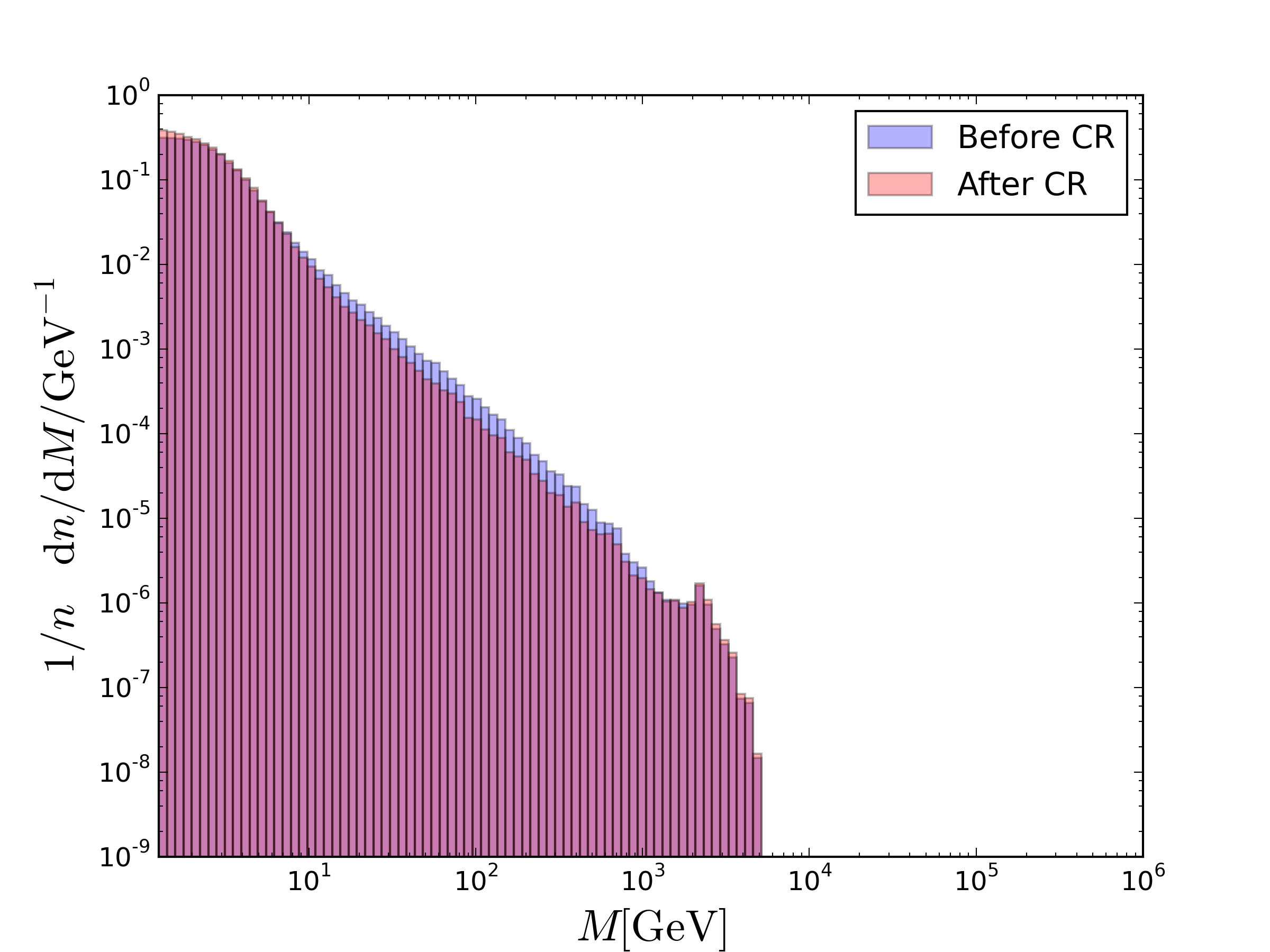}
\caption{Comparison between the distributions of invariant cluster masses 
before and after the colour reconnection.}
\label{fig:new}
\end{figure}

\begin{figure}[t]
\centering
\includegraphics[width=8cm]{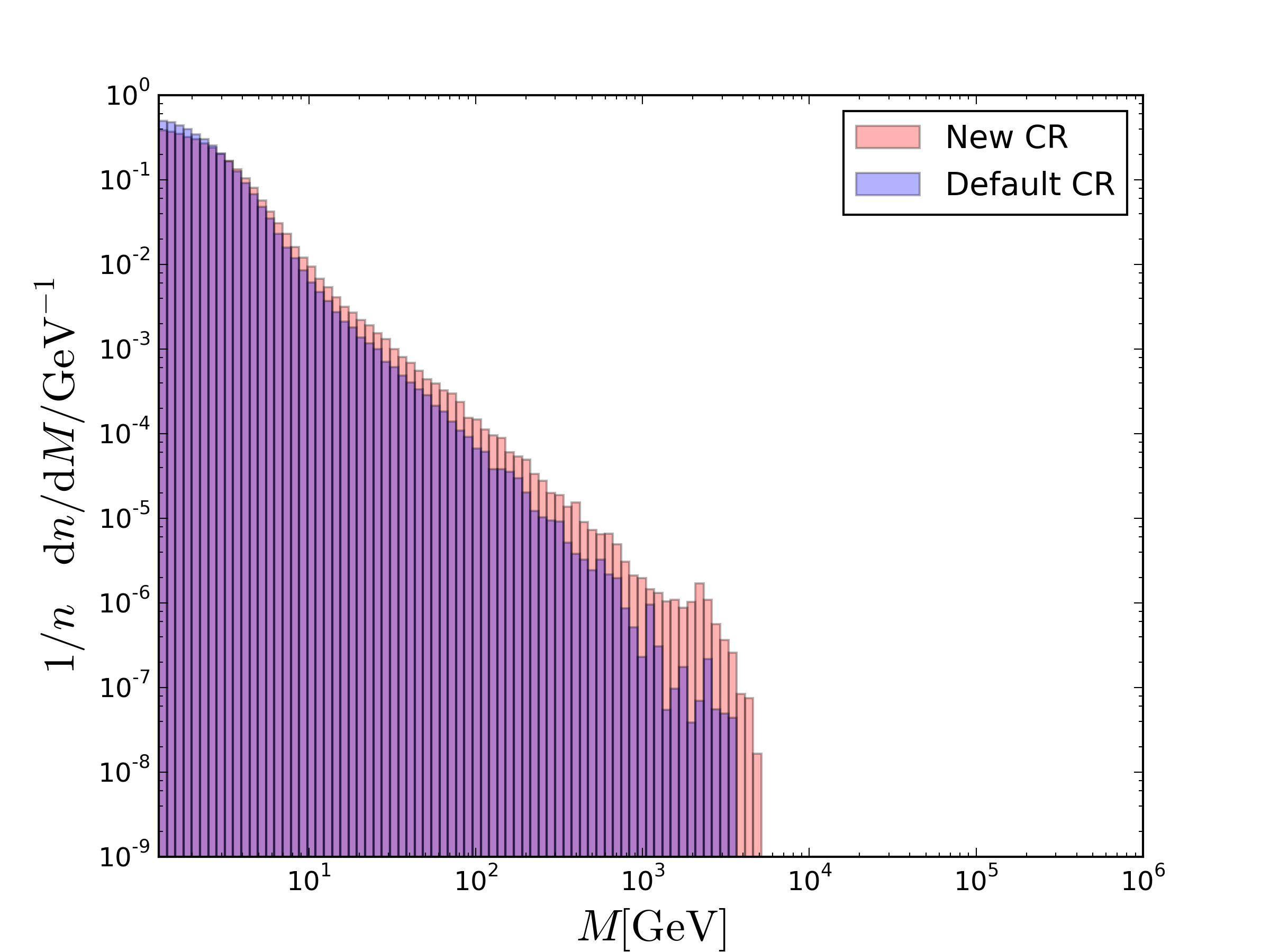}
\caption{Comparison between the distribution of invariant cluster masses after colour reconnection
for the old model and the new model.}
\label{fig:oldnew}
\end{figure}

\begin{figure}[t]
\centering
\includegraphics[width=8cm]{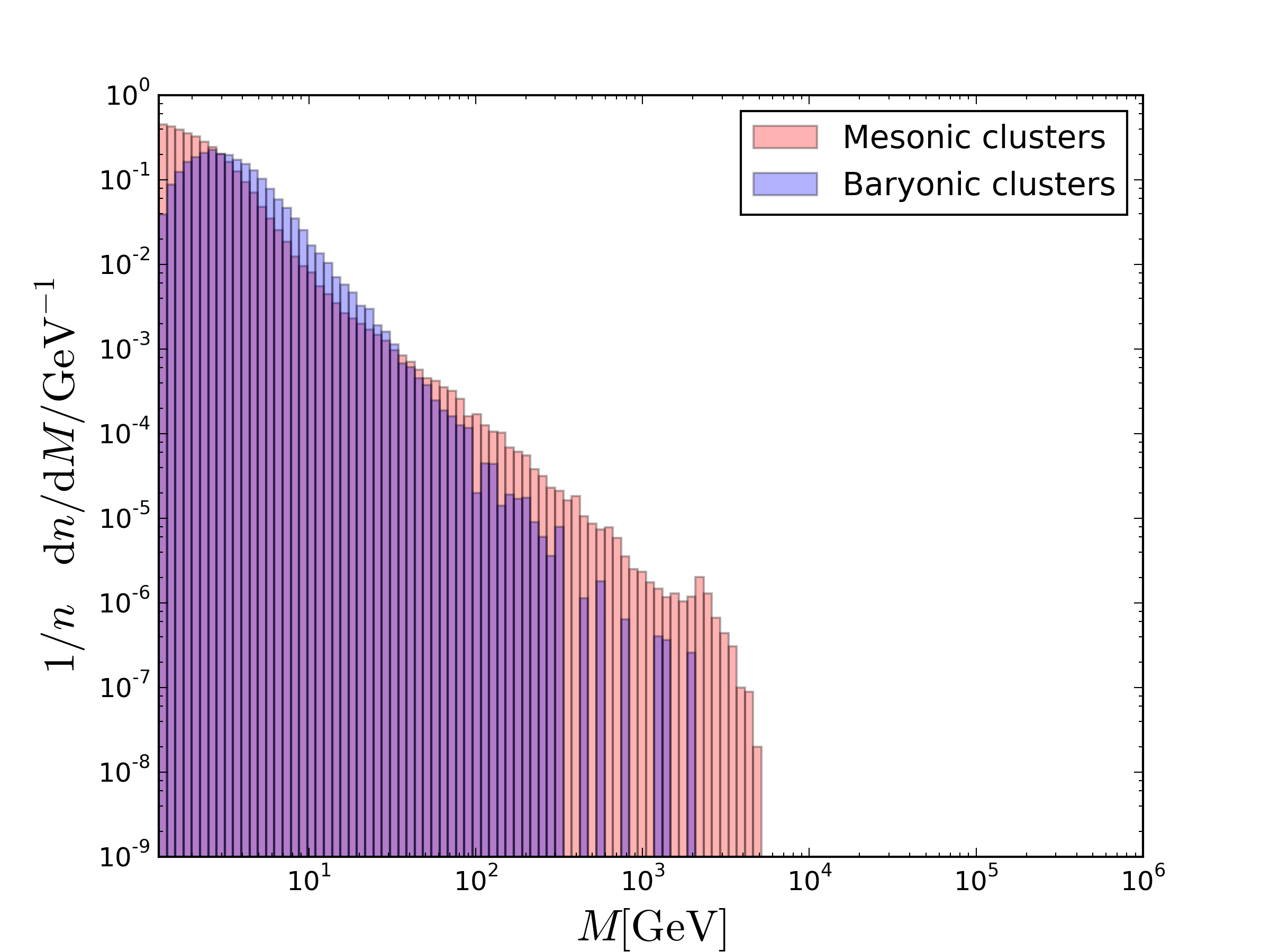}
\caption{The invariant mass distribution of mesonic and baryonic clusters after colour reconnection.}
\label{fig:baryonicmesonic}
\end{figure}

In this section we discuss the effects of the new model for colour
reconnection on the clusters and the distribution of cluster masses. This is
only done for non-diffractive events, since colour reconnection has no effect
on the simulation of diffraction.  In Fig.\,\ref{fig:new} we show the effects
of the new colour reconnection model on the distribution of cluster masses. It
can still be seen that after colour reconnection the cluster masses get
shifted towards smaller values as it was the case in the old model but the
effect is not as severe as in the old model (see Fig.\,\ref{fig:oldnew}).  In
a direct comparison between the default model and the new model we see that
the new model favours the production of more heavier clusters. In
Fig.\,\ref{fig:baryonicmesonic} we show the distribution of cluster masses
after colour reconnection separately for baryonic and mesonic clusters.  The
contribution in the high-mass region mainly comes from mesonic
clusters. Baryonic clusters dominate the mid-mass region between 1 and 13 GeV
while large baryonic clusters are highly suppressed.
With this in mind a picture of the cluster configuration emerges which, in
order to be able to describe the data, favours the production of baryonic
clusters with an intermediate cluster mass and small fluctuations towards
clusters of very high masses. In general one can say that smaller clusters
lead to less heavier particles due to the highly restricted phase space in the
cluster decay stage.  This makes the old default model for colour
reconnection, which is based on the reduction of cluster masses not able to
reproduce the observables concerned with heavier particles as discussed in
Sec.\,\ref{sec:results} due to a lack of heavy clusters.

\section{Conclusion and outlook}
\label{sec:conclusion}

We have implemented a new model for colour reconnection which is entirely
based on a geometrical picture instead of an algorithm that tries to directly
minimize the invariant cluster mass.  In addition we allow reconnections
between multiple mesonic clusters to form baryonic clusters which was not
possible in the old model.  With this mechanism we get an important lever on
the baryon to meson ratio which is a necessary starting point in order to
describe flavour observables.  The amount of reconnection also depends on the
multiplicity of the events which can be seen by comparing the model to the
charged particle multiplicities which get significantly better.  In addition
we allow for non-perturbative gluon splitting into strange quark-antiquark
pairs. Only with this additional source of strangeness it is possible to get a
good description of the $p_{\perp}$ spectra of the kaons.  The description of
the heavy baryons $\mathrm{\Lambda}$ and $\mathrm{\Xi^-}$ improves once we
combine the new model for colour reconnection and the additional source of
strangeness.  The model was tuned to 7\,TeV MB data and various hadron flavour
observables.  With the new model the full range of MB data can be described
with a similar good quality as the old model and in addition we improve the
description of hadron flavour observables significantly.

A comparison with ALICE data concerning the enhancement of (multi-)strange
baryons led us to the conclusion that our simple model is able to reproduce
the general trend of some of the observables, but fails to describe the data
in its entirety.  Nonetheless we see indications that the increase in the
strange baryon fraction can also be explained by an approach with colour
reconnection in combination with the cluster model. In order to arrive at a
satisfying description of high multiplicity events we suggest a model were
these events are originating from from less and heavier clusters instead of
many lighter ones.  This is necessary to raise the probability for producing
heavy and strangeness containing hadrons. This could also in principle be
solved in the framework of a colour reconnection model were many overlapping
clusters (or a region of high cluster density) fuse together to form a heavy
cluster which opens up the phase space for the production of strangeness and
baryons in the cluster decay stage.

This issue could possibly be addressed by a space-time picture of
cluster-evolution which will be left for future work.

A shortcoming of the model lies in the algorithm which is biased by the order
of clusters which are considered for reconnection and the fact that baryonic
clusters cannot be re-reconnected.  This will ultimately yield clusters which
do not consist of the nearest neighbours in phase space but a small overlap
between the clusters will still be present. Due to the sheer amount of
possibilities on how to assign baryonic clusters we are forced to introduce
some sort of arbitrariness when it comes to the cluster assignment. When
comparing the new model with the old model, we see that the new colour
reconnection model does not have the same effect on the invariant mass
distribution in terms of reduction of cluster masses but fuses mesonic
clusters together in order to form baryonic clusters and therefore adds an
additional possibility to produce heavy baryons. According to the data a
significant reduction in cluster mass is not favoured. The data prefers more
fluctuations in cluster size and explicitly welcomes the possibility to
produce baryonic clusters. Otherwise the production of heavy strange baryons
is not possible and highly suppressed.

Understanding soft physics remains difficult but new approaches and models are
necessary in order to improve the quality of Monte-Carlo event
generators. Overall, we have shown that small changes in the model for colour
reconnection and gluon-splitting mechanism can have significant effects on
some observables.

\section*{Acknowledgments}

We are grateful to the other members of the Herwig collaboration for critical
discussions and support. We would also like to thank Christian Bierlich and
Christian Holm Christensen for providing us with the ALICE data and the
analysis.  This work has received funding from the European Union's Horizon
2020 research and innovation programme as part of the Marie Skłodowska-Curie
Innovative Training Network MCnetITN3 (grant agreement no. 722104).  This work
has been supported in part by the BMBF under grant number 05H15VKCCA.  SP
acknowledges the kind hospitality of the Erwin Schr\"odinger Institute and the
Particle Physics group at the University of Vienna while part of this work has
been completed.

\bibliography{reco}
\begin{appendices}
\begin{table*}[htpb]
\centering

\begin{tabular}{llllr}
      & $p_{\perp,0}^{min}/\rm{GeV}$ & $\mu^2/\,{\rm{GeV}^2}$ & $p_{\rm{R}}$ & $p_{\rm{B}}$ \\
\midrule
default & 3.502 & 1.402 & 0.5    &  0\\
tune    & 3.269 & 1.963 & 0.543  &  0.2086 \\
\end{tabular}
\caption { Results of the parameter values from the tuning procedure that
  resulted in the smallest $\chi^2 / N_{\mathrm{dof}}$ value for $\sqrt{s}=7\,
  \mathrm{TeV}$ centre-of-mass energy compared with the default tune from
  Herwig 7.1.}
\label{table1}
\end{table*}

\begin{table*}[htpb]
\centering

\begin{tabular}{llllllr}
      & $p_{\perp,0}^{min}/\rm{GeV}$ & $\mu^2/\,{\rm{GeV}^2}$ & $p_{\rm{R}}$ & $p_{\rm{B}}$  &  PwtSquark & SplitPwtSquark \\
\midrule
default & 3.502 & 1.402 & 0.5    &  0     & 0.665  & 0     \\
tune    & 3.053 & 1.282 & 0.772  &  0.477 & 0.291  & 0.824 \\
\end{tabular}
\caption { Results of the parameter values from the tuning procedure that
  resulted in the smallest $\chi^2 / N_{\mathrm{dof}}$ value for $\sqrt{s}=7\,
  \mathrm{TeV}$ centre-of-mass energy compared with the default tune from
  Herwig 7.1.}
\label{table2}
\end{table*}

\end{appendices}
\end{document}